\documentclass[longauth]{aa_mod}

\usepackage{graphicx}
\usepackage{txfonts}
\usepackage[pdfencoding=auto,psdextra]{hyperref}
\hypersetup{
    colorlinks=true,
    linkcolor=blue,
    filecolor=magenta,      
    urlcolor=blue,
    citecolor=blue
}
\usepackage{lipsum}
\usepackage{microtype}
\usepackage{xcolor}
\usepackage{lscape}             
\usepackage{placeins}           
\usepackage{euclid}
\usepackage{booktabs, caption}
\usepackage{euclid}
\usepackage{orcidlink}
\let\orcid\orcidlink

\newcommand{\msun}{\ensuremath{M_\odot\, h^{-1}_{70} }}
\newcommand{\mass}{\mbox{$M_{\mbox{\scriptsize 200c}}$}}

\newcommand{\muk}{\ensuremath{\mu {\rm K_{CMB}}}}

\newcommand{\zcluster}{$2.92$}
\newcommand{\msz}{\mass  = ($2.1\pm0.25) \times 10^{14}$ \msun}

\titlerunning{SPT-CL~J0111.8$-$5637.6} 
\authorrunning{Bleem et al.}

\begin{document}

   \title{{SPT-3G and \Euclid: SPT-CL~J0111.8$-$5637.6 -- a massive SZ-selected galaxy cluster at $z=$ 2.92}\thanks{This paper is published on behalf of the SPT-3G and Euclid Collaborations.}}

%
%
%

\author{%
L.~E.~Bleem\orcid{0000-0001-7665-5079}\thanks{lbleem@anl.gov}\inst{\ref{aff1},\ref{aff2},\ref{aff3}}
\and
A.~H.~Gonzalez\orcid{0000-0002-0933-8601}\inst{\ref{aff4}}
\and
M.~McDonald\inst{\ref{aff5},\ref{aff6}}
\and
S.~A.~Stanford\orcid{0000-0003-0122-0841}\inst{\ref{aff7}}
\and
E.~Coleman\orcid{0009-0008-3927-4638}\inst{\ref{aff3}}
\and
M.~Klein\orcid{0000-0002-8248-4488}\inst{\ref{aff8}}
\and
M.~Archipley\orcid{0000-0002-0517-9842}\inst{\ref{aff3},\ref{aff2}}
\and
N.~Sulzenauer\orcid{0000-0002-3187-1648}\inst{\ref{aff9}}
\and
M.~D.~Gladders\orcid{0000-0003-1370-5010}\inst{\ref{aff3},\ref{aff2}}
\and
C.~Benoist\orcid{0009-0003-0880-6295}\inst{\ref{aff10}}
\and
S.~Hughes\orcid{0000-0002-7332-2751}\inst{\ref{aff5}}
\and
S.~Maurogordato\inst{\ref{aff10}}
\and
S.~Andreon\orcid{0000-0002-2041-8784}\inst{\ref{aff11}}
\and
S.~W.~Allen\orcid{0000-0003-0667-5941}\inst{\ref{aff12},\ref{aff13},\ref{aff14}}
\and
B.~Altieri\orcid{0000-0003-3936-0284}\inst{\ref{aff15}}
\and
D.~Anbajagane\inst{\ref{aff3},\ref{aff2}}
\and
E.~Anderes\orcid{0009-0003-3245-3979}\inst{\ref{aff16}}
\and
A.~J.~Anderson\orcid{0000-0002-4435-4623}\inst{\ref{aff17},\ref{aff2},\ref{aff3}}
\and
B.~Ansarinejad\inst{\ref{aff18}}
\and
M.~Aravena\inst{\ref{aff19},\ref{aff20}}
\and
N.~Auricchio\orcid{0000-0003-4444-8651}\inst{\ref{aff21}}
\and
C.~Baccigalupi\orcid{0000-0002-8211-1630}\inst{\ref{aff22},\ref{aff23},\ref{aff24},\ref{aff25}}
\and
M.~Baldi\orcid{0000-0003-4145-1943}\inst{\ref{aff26},\ref{aff21},\ref{aff27}}
\and
L.~Balkenhol\orcid{0000-0001-6899-1873}\inst{\ref{aff28}}
\and
S.~Bardelli\orcid{0000-0002-8900-0298}\inst{\ref{aff21}}
\and
D.~R.~Barron\orcid{0000-0002-1623-5651}\inst{\ref{aff29}}
\and
P.~S.~Barry\inst{\ref{aff30}}
\and
J.~G.~Bartlett\orcid{0000-0002-0685-8310}\inst{\ref{aff31},\ref{aff32}}
\and
P.~Battaglia\orcid{0000-0002-7337-5909}\inst{\ref{aff21}}
\and
L.~Baumont\orcid{0000-0002-1518-0150}\inst{\ref{aff33},\ref{aff23},\ref{aff22}}
\and
M.~Bayliss\inst{\ref{aff34}}
\and
K.~Benabed\inst{\ref{aff28}}
\and
A.~N.~Bender\orcid{0000-0001-5868-0748}\inst{\ref{aff1},\ref{aff2},\ref{aff3}}
\and
B.~A.~Benson\orcid{0000-0002-5108-6823}\inst{\ref{aff17},\ref{aff2},\ref{aff3}}
\and
D.~Berard\orcid{0000-0001-7654-6809}\inst{\ref{aff10}}
\and
F.~Bianchini\orcid{0000-0003-4847-3483}\inst{\ref{aff12},\ref{aff13},\ref{aff14}}
\and
A.~Biviano\orcid{0000-0002-0857-0732}\inst{\ref{aff23},\ref{aff22}}
\and
S.~Bocquet\orcid{0000-0002-4900-805X}\inst{\ref{aff8}}
\and
M.~Bolzonella\orcid{0000-0003-3278-4607}\inst{\ref{aff21}}
\and
S.~Borgani\orcid{0000-0001-6151-6439}\inst{\ref{aff33},\ref{aff22},\ref{aff23},\ref{aff24},\ref{aff35}}
\and
F.~R.~Bouchet\orcid{0000-0002-8051-2924}\inst{\ref{aff28}}
\and
E.~Branchini\orcid{0000-0002-0808-6908}\inst{\ref{aff36},\ref{aff37},\ref{aff11}}
\and
M.~Brescia\orcid{0000-0001-9506-5680}\inst{\ref{aff38},\ref{aff39}}
\and
M.~Calzadilla\inst{\ref{aff40}}
\and
S.~Camera\orcid{0000-0003-3399-3574}\inst{\ref{aff41},\ref{aff42},\ref{aff43}}
\and
E.~Camphuis\orcid{0000-0003-3483-8461}\inst{\ref{aff28}}
\and
M.~G.~Campitiello\inst{\ref{aff1}}
\and
V.~Capobianco\orcid{0000-0002-3309-7692}\inst{\ref{aff43}}
\and
A.~Cappi\inst{\ref{aff10},\ref{aff21}}
\and
C.~Carbone\orcid{0000-0003-0125-3563}\inst{\ref{aff44}}
\and
J.~E.~Carlstrom\orcid{0000-0002-2044-7665}\inst{\ref{aff2},\ref{aff45},\ref{aff46},\ref{aff1},\ref{aff3}}
\and
J.~Carretero\orcid{0000-0002-3130-0204}\inst{\ref{aff47},\ref{aff48}}
\and
J.~Carron\orcid{0000-0002-5751-1392}\inst{\ref{aff49},\ref{aff50}}
\and
M.~Castellano\orcid{0000-0001-9875-8263}\inst{\ref{aff51}}
\and
G.~Castignani\orcid{0000-0001-6831-0687}\inst{\ref{aff21}}
\and
S.~Cavuoti\orcid{0000-0002-3787-4196}\inst{\ref{aff39},\ref{aff52}}
\and
C.~L.~Chang\inst{\ref{aff1},\ref{aff2},\ref{aff3}}
\and
S.~C.~Chapman\inst{\ref{aff53},\ref{aff54},\ref{aff55}}
\and
P.~M.~Chichura\orcid{0000-0002-5397-9035}\inst{\ref{aff46},\ref{aff2}}
\and
A.~Chokshi\inst{\ref{aff3}}
\and
T.-L.~Chou\orcid{0000-0002-3091-8790}\inst{\ref{aff3},\ref{aff2},\ref{aff56}}
\and
A.~Cimatti\inst{\ref{aff57}}
\and
A.~Coerver\orcid{0000-0002-2707-1672}\inst{\ref{aff58}}
\and
C.~Colodro-Conde\inst{\ref{aff59}}
\and
G.~Congedo\orcid{0000-0003-2508-0046}\inst{\ref{aff60}}
\and
C.~J.~Conselice\orcid{0000-0003-1949-7638}\inst{\ref{aff61}}
\and
L.~Conversi\orcid{0000-0002-6710-8476}\inst{\ref{aff62},\ref{aff15}}
\and
Y.~Copin\orcid{0000-0002-5317-7518}\inst{\ref{aff63}}
\and
F.~Courbin\orcid{0000-0003-0758-6510}\inst{\ref{aff64},\ref{aff65},\ref{aff66}}
\and
H.~M.~Courtois\orcid{0000-0003-0509-1776}\inst{\ref{aff67}}
\and
T.~M.~Crawford\orcid{0000-0001-9000-5013}\inst{\ref{aff3},\ref{aff2}}
\and
M.~Cropper\orcid{0000-0003-4571-9468}\inst{\ref{aff68}}
\and
C.~Daley\orcid{0000-0002-3760-2086}\inst{\ref{aff69},\ref{aff70}}
\and
S.~Damish\inst{\ref{aff46},\ref{aff2}}
\and
H.~Dannerbauer\orcid{0000-0001-7147-3575}\inst{\ref{aff59},\ref{aff71}}
\and
T.~de~Haan\orcid{0000-0001-5105-9473}\inst{\ref{aff72}}
\and
G.~De~Lucia\orcid{0000-0002-6220-9104}\inst{\ref{aff23}}
\and
R.~P.~Deane\inst{\ref{aff73},\ref{aff74},\ref{aff75}}
\and
H.~Degaudenzi\orcid{0000-0002-5887-6799}\inst{\ref{aff76}}
\and
L.~Di~Mascolo\inst{\ref{aff77}}
\and
K.~R.~Dibert\inst{\ref{aff3},\ref{aff2}}
\and
M.~A.~Dobbs\inst{\ref{aff78},\ref{aff79}}
\and
H.~Dole\orcid{0000-0002-9767-3839}\inst{\ref{aff80},\ref{aff81}}
\and
M.~Doohan\inst{\ref{aff18}}
\and
F.~Dubath\orcid{0000-0002-6533-2810}\inst{\ref{aff76}}
\and
X.~Dupac\inst{\ref{aff15}}
\and
D.~Dutcher\orcid{0000-0002-9962-2058}\inst{\ref{aff82}}
\and
A.~Ealet\orcid{0000-0003-3070-014X}\inst{\ref{aff63}}
\and
P.~R.~M.~Eisenhardt\inst{\ref{aff83}}
\and
G.~Fabbian\orcid{0000-0002-3255-4695}\inst{\ref{aff80}}
\and
M.~Fabricius\orcid{0000-0002-7025-6058}\inst{\ref{aff84},\ref{aff85}}
\and
M.~Farina\orcid{0000-0002-3089-7846}\inst{\ref{aff86}}
\and
F.~Faustini\orcid{0000-0001-6274-5145}\inst{\ref{aff51}}
\and
C.~Feng\inst{\ref{aff87},\ref{aff88},\ref{aff89}}
\and
K.~R.~Ferguson\orcid{0000-0002-4928-8813}\inst{\ref{aff90},\ref{aff91}}
\and
N.~C.~Ferree\orcid{0000-0002-7130-7099}\inst{\ref{aff92},\ref{aff12},\ref{aff13}}
\and
S.~Ferriol\inst{\ref{aff63}}
\and
K.~Fichman\inst{\ref{aff46},\ref{aff2}}
\and
F.~Finelli\orcid{0000-0002-6694-3269}\inst{\ref{aff21},\ref{aff93}}
\and
K.~Finner\orcid{0000-0002-4462-0709}\inst{\ref{aff94}}
\and
B.~Floyd\inst{\ref{aff95}}
\and
E.~Foley\inst{\ref{aff55}}
\and
K.~Fosdick\inst{\ref{aff6}}
\and
A.~Foster\orcid{0000-0002-7145-1824}\inst{\ref{aff82}}
\and
N.~Fourmanoit\orcid{0009-0005-6816-6925}\inst{\ref{aff96}}
\and
M.~Fumana\orcid{0000-0001-6787-5950}\inst{\ref{aff44}}
\and
L.~Gabarra\orcid{0000-0002-8486-8856}\inst{\ref{aff97}}
\and
S.~Galeotta\orcid{0000-0002-3748-5115}\inst{\ref{aff23}}
\and
S.~Galli\inst{\ref{aff28}}
\and
A.~E.~Gambrel\inst{\ref{aff2}}
\and
A.~K.~Gao\inst{\ref{aff89}}
\and
J.~Garc\'ia-Bellido\orcid{0000-0002-9370-8360}\inst{\ref{aff98}}
\and
F.~Ge\orcid{0000-0002-3833-8133}\inst{\ref{aff92},\ref{aff12},\ref{aff13},\ref{aff7}}
\and
K.~George\orcid{0000-0002-1734-8455}\inst{\ref{aff8}}
\and
Z.~Ghaffari\orcid{0000-0002-6467-8078}\inst{\ref{aff23},\ref{aff22}}
\and
C.~Giocoli\orcid{0000-0002-9590-7961}\inst{\ref{aff21},\ref{aff27}}
\and
G.~Gozaliasl\orcid{0000-0002-0236-919X}\inst{\ref{aff99},\ref{aff100}}
\and
J.~Gracia-Carpio\orcid{0000-0003-4689-3134}\inst{\ref{aff84}}
\and
S.~Grandis\inst{\ref{aff101}}
\and
A.~Grazian\orcid{0000-0002-5688-0663}\inst{\ref{aff102}}
\and
F.~Grupp\inst{\ref{aff84},\ref{aff85}}
\and
F.~Guidi\orcid{0000-0001-7593-3962}\inst{\ref{aff7},\ref{aff28}}
\and
S.~Guns\inst{\ref{aff58}}
\and
C.~M.~Gutierrez\orcid{0000-0001-7854-783X}\inst{\ref{aff59},\ref{aff71}}
\and
L.~Guzzo\orcid{0000-0001-8264-5192}\inst{\ref{aff103},\ref{aff11},\ref{aff104}}
\and
N.~W.~Halverson\inst{\ref{aff105},\ref{aff106}}
\and
S.~V.~H.~Haugan\orcid{0000-0001-9648-7260}\inst{\ref{aff107}}
\and
S.~Hemmati\orcid{0000-0003-2226-5395}\inst{\ref{aff94}}
\and
R.~Hill\inst{\ref{aff53}}
\and
E.~Hivon\orcid{0000-0003-1880-2733}\inst{\ref{aff28}}
\and
H.~Hoekstra\orcid{0000-0002-0641-3231}\inst{\ref{aff108}}
\and
G.~P.~Holder\orcid{0000-0002-0463-6394}\inst{\ref{aff89}}
\and
W.~L.~Holzapfel\orcid{0000-0002-5744-6439}\inst{\ref{aff58}}
\and
J.~C.~Hood\inst{\ref{aff2}}
\and
F.~Hormuth\inst{\ref{aff109}}
\and
A.~Hornstrup\orcid{0000-0002-3363-0936}\inst{\ref{aff110},\ref{aff111}}
\and
A.~Hryciuk\inst{\ref{aff46},\ref{aff2}}
\and
N.~Huang\orcid{0000-0003-3595-0359}\inst{\ref{aff58}}
\and
R.~Ingrao\orcid{0009-0009-6082-9818}\inst{\ref{aff33},\ref{aff22},\ref{aff23},\ref{aff24}}
\and
K.~Jahnke\orcid{0000-0003-3804-2137}\inst{\ref{aff112}}
\and
M.~Jauzac\orcid{0000-0003-1974-8732}\inst{\ref{aff113},\ref{aff114},\ref{aff115},\ref{aff116},\ref{aff117}}
\and
M.~Jhabvala\inst{\ref{aff118}}
\and
T.~Jhaveri\inst{\ref{aff3},\ref{aff2}}
\and
S.~Kermiche\orcid{0000-0002-0302-5735}\inst{\ref{aff96}}
\and
F.~K\'eruzor\'e\inst{\ref{aff1}}
\and
A.~R.~Khalife\orcid{0000-0002-8388-4950}\inst{\ref{aff28}}
\and
G.~Khullar\inst{\ref{aff119},\ref{aff120}}
\and
K.~Kim\orcid{0000-0001-6505-0293}\inst{\ref{aff94}}
\and
L.~Knox\inst{\ref{aff7}}
\and
K.~Kornoelje\inst{\ref{aff3},\ref{aff2},\ref{aff1}}
\and
B.~Kubik\orcid{0009-0006-5823-4880}\inst{\ref{aff63}}
\and
M.~K\"ummel\orcid{0000-0003-2791-2117}\inst{\ref{aff85}}
\and
M.~Kunz\orcid{0000-0002-3052-7394}\inst{\ref{aff121}}
\and
C.-L.~Kuo\inst{\ref{aff12},\ref{aff13},\ref{aff14}}
\and
H.~Kurki-Suonio\orcid{0000-0002-4618-3063}\inst{\ref{aff100},\ref{aff122}}
\and
Q.~Le~Boulc'h\inst{\ref{aff123}}
\and
A.~M.~C.~Le~Brun\orcid{0000-0002-0936-4594}\inst{\ref{aff124}}
\and
V.~Le~Brun\orcid{0000-0002-5027-1939}\inst{\ref{aff125}}
\and
G.~Leroy\orcid{0009-0004-2523-4425}\inst{\ref{aff113},\ref{aff114}}
\and
N.~LeVar\inst{\ref{aff126}}
\and
K.~Levy\inst{\ref{aff18}}
\and
Y.~Li\orcid{0000-0002-4820-1122}\inst{\ref{aff2}}
\and
S.~Ligori\orcid{0000-0003-4172-4606}\inst{\ref{aff43}}
\and
P.~B.~Lilje\orcid{0000-0003-4324-7794}\inst{\ref{aff107}}
\and
V.~Lindholm\orcid{0000-0003-2317-5471}\inst{\ref{aff100},\ref{aff122}}
\and
I.~Lloro\orcid{0000-0001-5966-1434}\inst{\ref{aff127}}
\and
A.~E.~Lowitz\orcid{0000-0002-4747-4276}\inst{\ref{aff2}}
\and
C.~Lu\inst{\ref{aff89}}
\and
G.~P.~Lynch\orcid{0009-0004-3143-1708}\inst{\ref{aff7}}
\and
T.~J.~Maccarone\orcid{0000-0003-0976-4755}\inst{\ref{aff128}}
\and
J.~F.~Macias-Perez\orcid{0000-0002-5385-2763}\inst{\ref{aff129}}
\and
M.~Magliocchetti\orcid{0000-0001-9158-4838}\inst{\ref{aff86}}
\and
G.~Mahler\inst{\ref{aff130}}
\and
G.~Mainetti\orcid{0000-0003-2384-2377}\inst{\ref{aff123}}
\and
M.~Malkan\inst{\ref{aff90}}
\and
A.~S.~Maniyar\orcid{0000-0002-4617-9320}\inst{\ref{aff131},\ref{aff13}}
\and
A.~Mantz\orcid{0000-0002-8031-1217}\inst{\ref{aff12}}
\and
O.~Marggraf\orcid{0000-0001-7242-3852}\inst{\ref{aff132}}
\and
N.~Martinet\orcid{0000-0003-2786-7790}\inst{\ref{aff125}}
\and
E.~S.~Martsen\orcid{0009-0009-0719-9781}\inst{\ref{aff3},\ref{aff2}}
\and
F.~Marulli\orcid{0000-0002-8850-0303}\inst{\ref{aff133},\ref{aff21},\ref{aff27}}
\and
R.~J.~Massey\orcid{0000-0002-6085-3780}\inst{\ref{aff114}}
\and
M.~Maturi\orcid{0000-0002-3517-2422}\inst{\ref{aff134},\ref{aff135}}
\and
B.~J.~Maughan\orcid{0000-0003-0791-9098}\inst{\ref{aff136}}
\and
E.~Medinaceli\orcid{0000-0002-4040-7783}\inst{\ref{aff21}}
\and
S.~Mei\orcid{0000-0002-2849-559X}\inst{\ref{aff31},\ref{aff32}}
\and
J.-B.~Melin\orcid{0009-0007-8652-1657}\inst{\ref{aff137}}
\and
O.~Melnychenko\orcid{0009-0000-5234-6896}\inst{\ref{aff40}}
\and
F.~Menanteau\inst{\ref{aff138},\ref{aff70},\ref{aff139}}
\and
M.~Meneghetti\orcid{0000-0003-1225-7084}\inst{\ref{aff21},\ref{aff27}}
\and
E.~Merlin\orcid{0000-0001-6870-8900}\inst{\ref{aff102}}
\and
G.~Meylan\orcid{0000-0001-6503-0209}\inst{\ref{aff140}}
\and
K.~Migkas\orcid{0000-0003-0451-0449}\inst{\ref{aff108}}
\and
M.~Millea\orcid{0000-0001-7317-0551}\inst{\ref{aff58}}
\and
J.~J.~Mohr\orcid{0000-0002-6875-2087}\inst{\ref{aff8}}
\and
P.~Monaco\orcid{0000-0003-2083-7564}\inst{\ref{aff33},\ref{aff23},\ref{aff24},\ref{aff22}}
\and
J.~Montgomery\inst{\ref{aff78}}
\and
A.~Mora\orcid{0000-0002-1922-8529}\inst{\ref{aff141}}
\and
M.~Moresco\orcid{0000-0002-7616-7136}\inst{\ref{aff133},\ref{aff21},\ref{aff27}}
\and
C.~Moretti\orcid{0000-0003-3314-8936}\inst{\ref{aff23},\ref{aff22},\ref{aff24}}
\and
L.~Moscardini\orcid{0000-0002-3473-6716}\inst{\ref{aff133},\ref{aff21},\ref{aff27}}
\and
E.~Munari\orcid{0000-0002-1751-5946}\inst{\ref{aff23},\ref{aff22}}
\and
Y.~Nakato\inst{\ref{aff13}}
\and
T.~Natoli\inst{\ref{aff2}}
\and
R.~C.~Nichol\orcid{0000-0003-0939-6518}\inst{\ref{aff142}}
\and
S.-M.~Niemi\orcid{0009-0005-0247-0086}\inst{\ref{aff143}}
\and
J.~W.~Nightingale\orcid{0000-0002-8987-7401}\inst{\ref{aff144}}
\and
V.~Nistane\orcid{0000-0002-2165-1393}\inst{\ref{aff10}}
\and
Y.~Omori\orcid{0000-0002-0963-7310}\inst{\ref{aff3},\ref{aff2},\ref{aff145}}
\and
A.~Ouellette\orcid{0000-0003-0170-5638}\inst{\ref{aff89}}
\and
F.~Pacaud\orcid{0000-0002-6622-4555}\inst{\ref{aff132},\ref{aff146}}
\and
C.~Padilla\orcid{0000-0001-7951-0166}\inst{\ref{aff147}}
\and
S.~Paltani\orcid{0000-0002-8108-9179}\inst{\ref{aff76}}
\and
Z.~Pan\orcid{0000-0002-6164-9861}\inst{\ref{aff1},\ref{aff2},\ref{aff46}}
\and
M.~Pannella\inst{\ref{aff23},\ref{aff22}}
\and
F.~Pasian\orcid{0000-0002-4869-3227}\inst{\ref{aff23}}
\and
W.~J.~Percival\orcid{0000-0002-0644-5727}\inst{\ref{aff148},\ref{aff149},\ref{aff150}}
\and
V.~Pettorino\orcid{0000-0002-4203-9320}\inst{\ref{aff143}}
\and
K.~A.~Phadke\orcid{0000-0001-7946-557X}\inst{\ref{aff70},\ref{aff139},\ref{aff145}}
\and
S.~Pires\orcid{0000-0002-0249-2104}\inst{\ref{aff69}}
\and
G.~Polenta\orcid{0000-0003-4067-9196}\inst{\ref{aff151}}
\and
M.~Poncet\inst{\ref{aff152}}
\and
L.~A.~Popa\inst{\ref{aff153}}
\and
K.~Prabhu\inst{\ref{aff7}}
\and
W.~Quan\orcid{0009-0002-2589-5501}\inst{\ref{aff1},\ref{aff46},\ref{aff2}}
\and
S.~Raghunathan\orcid{0000-0003-1405-378X}\inst{\ref{aff7},\ref{aff139}}
\and
M.~Rahimi\inst{\ref{aff18}}
\and
A.~Rahlin\orcid{0000-0003-3953-1776}\inst{\ref{aff3},\ref{aff2}}
\and
F.~Raison\orcid{0000-0002-7819-6918}\inst{\ref{aff84}}
\and
C.~L.~Reichardt\orcid{0000-0003-2226-9169}\inst{\ref{aff18}}
\and
A.~Renzi\orcid{0000-0001-9856-1970}\inst{\ref{aff154},\ref{aff155}}
\and
J.~Rhodes\orcid{0000-0002-4485-8549}\inst{\ref{aff83}}
\and
G.~Riccio\inst{\ref{aff39}}
\and
I.~Risso\orcid{0000-0003-2525-7761}\inst{\ref{aff36},\ref{aff37},\ref{aff11}}
\and
F.~Rizzo\orcid{0000-0002-9407-585X}\inst{\ref{aff23}}
\and
E.~Romelli\orcid{0000-0003-3069-9222}\inst{\ref{aff23}}
\and
M.~Roncarelli\orcid{0000-0001-9587-7822}\inst{\ref{aff21}}
\and
M.~Rouble\inst{\ref{aff78}}
\and
J.~E.~Ruhl\inst{\ref{aff156}}
\and
M.~Rybak\inst{\ref{aff60}}
\and
R.~Saglia\orcid{0000-0003-0378-7032}\inst{\ref{aff85},\ref{aff84}}
\and
Z.~Sakr\orcid{0000-0002-4823-3757}\inst{\ref{aff98},\ref{aff117},\ref{aff157}}
\and
D.~Sapone\orcid{0000-0001-7089-4503}\inst{\ref{aff158}}
\and
A.~Saro\inst{\ref{aff33},\ref{aff22}}
\and
B.~Sartoris\orcid{0000-0003-1337-5269}\inst{\ref{aff10}}
\and
M.~Schirmer\orcid{0000-0003-2568-9994}\inst{\ref{aff112}}
\and
P.~Schneider\orcid{0000-0001-8561-2679}\inst{\ref{aff132}}
\and
T.~Schrabback\orcid{0000-0002-6987-7834}\inst{\ref{aff101}}
\and
D.~Scott\orcid{0000-0002-6878-9840}\inst{\ref{aff53}}
\and
A.~Secroun\orcid{0000-0003-0505-3710}\inst{\ref{aff96}}
\and
M.~Sereno\orcid{0000-0003-0302-0325}\inst{\ref{aff21},\ref{aff27}}
\and
K.~Sharon\orcid{0000-0002-7559-0864}\inst{\ref{aff159}}
\and
A.~C.~Silva~Oliveira\orcid{0000-0001-5755-5865}\inst{\ref{aff92},\ref{aff12},\ref{aff13}}
\and
P.~Simon\inst{\ref{aff132}}
\and
A.~Simpson\inst{\ref{aff3},\ref{aff2}}
\and
C.~Sirignano\orcid{0000-0002-0995-7146}\inst{\ref{aff154},\ref{aff155}}
\and
G.~Sirri\orcid{0000-0003-2626-2853}\inst{\ref{aff27}}
\and
J.~A.~Sobrin\orcid{0000-0001-6155-5315}\inst{\ref{aff160},\ref{aff17}}
\and
T.~Somboonpanyakul\inst{\ref{aff161}}
\and
J.~S.~Spilker\inst{\ref{aff126}}
\and
L.~Stanco\orcid{0000-0002-9706-5104}\inst{\ref{aff155}}
\and
A.~A.~Stark\inst{\ref{aff40}}
\and
D.~Stern\orcid{0000-0003-2686-9241}\inst{\ref{aff83}}
\and
V.~Strazzullo\inst{\ref{aff23},\ref{aff22}}
\and
P.~Tallada-Cresp\'{i}\orcid{0000-0002-1336-8328}\inst{\ref{aff47},\ref{aff48}}
\and
C.~Tandoi\orcid{0000-0002-2077-6004}\inst{\ref{aff70}}
\and
K.~Tanidis\orcid{0000-0001-9843-5130}\inst{\ref{aff162}}
\and
A.~N.~Taylor\inst{\ref{aff60}}
\and
Q.~Taylor\orcid{0000-0002-6829-4389}\inst{\ref{aff156},\ref{aff163}}
\and
I.~Tereno\orcid{0000-0002-4537-6218}\inst{\ref{aff164},\ref{aff165}}
\and
S.~Toft\orcid{0000-0003-3631-7176}\inst{\ref{aff166},\ref{aff167}}
\and
R.~Toledo-Moreo\orcid{0000-0002-2997-4859}\inst{\ref{aff168},\ref{aff169}}
\and
G.~Toni\orcid{0009-0005-3133-1157}\inst{\ref{aff133},\ref{aff21},\ref{aff135}}
\and
F.~Torradeflot\orcid{0000-0003-1160-1517}\inst{\ref{aff48},\ref{aff47}}
\and
C.~Trendafilova\orcid{0000-0001-5500-4058}\inst{\ref{aff163},\ref{aff139}}
\and
A.~Tsyganov\inst{\ref{aff170}}
\and
I.~Tutusaus\orcid{0000-0002-3199-0399}\inst{\ref{aff171},\ref{aff172},\ref{aff117}}
\and
T.~Vassallo\orcid{0000-0001-6512-6358}\inst{\ref{aff23},\ref{aff8}}
\and
J.~D.~Vieira\orcid{0000-0001-7192-3871}\inst{\ref{aff70},\ref{aff89},\ref{aff139}}
\and
A.~G.~Vieregg\orcid{0000-0002-4528-9886}\inst{\ref{aff2},\ref{aff3},\ref{aff45},\ref{aff46}}
\and
A.~Vitrier\orcid{0009-0009-3168-092X}\inst{\ref{aff28}}
\and
D.~Vizgan\inst{\ref{aff70},\ref{aff173}}
\and
Y.~Wan\orcid{0009-0007-0072-0445}\inst{\ref{aff70},\ref{aff139}}
\and
Y.~Wang\orcid{0000-0002-4749-2984}\inst{\ref{aff94}}
\and
A.~Wei\ss{}\inst{\ref{aff9}}
\and
J.~Weller\orcid{0000-0002-8282-2010}\inst{\ref{aff85},\ref{aff84}}
\and
N.~Whitehorn\orcid{0000-0002-3157-0407}\inst{\ref{aff91}}
\and
W.~L.~K.~Wu\orcid{0000-0001-5411-6920}\inst{\ref{aff92},\ref{aff12},\ref{aff14}}
\and
M.~R.~Young\inst{\ref{aff17},\ref{aff2}}
\and
G.~Zamorani\orcid{0000-0002-2318-301X}\inst{\ref{aff21}}
\and
J.~A.~Zebrowski\orcid{0000-0003-2375-0229}\inst{\ref{aff2},\ref{aff3},\ref{aff17}}
\and
F.~M.~Zerbi\orcid{0000-0002-9996-973X}\inst{\ref{aff11}}
\and
D.~Zhou\inst{\ref{aff53}}
\and
E.~Zucca\orcid{0000-0002-5845-8132}\inst{\ref{aff21}}
}

\institute{%
High-Energy Physics Division, Argonne National Laboratory, 9700 South Cass Avenue, Lemont, IL, 60439, USA\label{aff1}
\and
Kavli Institute for Cosmological Physics, University of Chicago, Chicago, IL 60637, USA\label{aff2}
\and
Department of Astronomy and Astrophysics, University of Chicago, 5640 South Ellis Avenue, Chicago, IL 60637, USA\label{aff3}
\and
Department of Astronomy, University of Florida, Bryant Space Science Center, Gainesville, FL 32611, USA\label{aff4}
\and
MIT Kavli Institute for Astrophysics and Space Research, Massachusetts Institute of Technology, Cambridge, MA 02139, USA\label{aff5}
\and
Department of Physics, Massachusetts Institute of Technology, 77 Massachusetts Avenue, Cambridge, MA 02139, USA\label{aff6}
\and
Department of Physics and Astronomy, University of California, Davis, CA 95616, USA\label{aff7}
\and
University Observatory, LMU Faculty of Physics, Scheinerstr.~1, 81679 Munich, Germany\label{aff8}
\and
Max-Planck-Institut f\"ur Radioastronomie, Auf dem H\"ugel 69, 53121 Bonn, Germany\label{aff9}
\and
Universit\'e C\^{o}te d'Azur, Observatoire de la C\^{o}te d'Azur, CNRS, Laboratoire Lagrange, Bd de l'Observatoire, CS 34229, 06304 Nice cedex 4, France\label{aff10}
\and
INAF-Osservatorio Astronomico di Brera, Via Brera 28, 20122 Milano, Italy\label{aff11}
\and
Kavli Institute for Particle Astrophysics and Cosmology, Stanford University, 452 Lomita Mall, Stanford, CA, 94305, USA\label{aff12}
\and
Department of Physics, Stanford University, 382 Via Pueblo Mall, Stanford, CA, 94305, USA\label{aff13}
\and
SLAC National Accelerator Laboratory, 2575 Sand Hill Road, Menlo Park, CA, 94025, USA\label{aff14}
\and
ESAC/ESA, Camino Bajo del Castillo, s/n., Urb. Villafranca del Castillo, 28692 Villanueva de la Ca\~nada, Madrid, Spain\label{aff15}
\and
Department of Statistics, University of California, One Shields Avenue, Davis, CA 95616, USA\label{aff16}
\and
Fermi National Accelerator Laboratory, MS209, P.O. Box 500, Batavia, IL, 60510, USA\label{aff17}
\and
School of Physics, University of Melbourne, Parkville, VIC 3010, Australia\label{aff18}
\and
Instituto de Estudios Astrof\'{\i}cos, Facultad de Ingenier\'{\i}a y Ciencias, Universidad Diego Portales, Av. Ej\'ercito 441, Santiago, Chile\label{aff19}
\and
Millennium Nucleus for Galaxies (MINGAL), Chile\label{aff20}
\and
INAF-Osservatorio di Astrofisica e Scienza dello Spazio di Bologna, Via Piero Gobetti 93/3, 40129 Bologna, Italy\label{aff21}
\and
IFPU, Institute for Fundamental Physics of the Universe, via Beirut 2, 34151 Trieste, Italy\label{aff22}
\and
INAF-Osservatorio Astronomico di Trieste, Via G. B. Tiepolo 11, 34143 Trieste, Italy\label{aff23}
\and
INFN, Sezione di Trieste, Via Valerio 2, 34127 Trieste TS, Italy\label{aff24}
\and
SISSA, International School for Advanced Studies, Via Bonomea 265, 34136 Trieste TS, Italy\label{aff25}
\and
Dipartimento di Fisica e Astronomia, Universit\`a di Bologna, Via Gobetti 93/2, 40129 Bologna, Italy\label{aff26}
\and
INFN-Sezione di Bologna, Viale Berti Pichat 6/2, 40127 Bologna, Italy\label{aff27}
\and
Sorbonne Universit\'e, CNRS, UMR 7095, Institut d'Astrophysique de Paris, 98 bis bd Arago, 75014 Paris, France\label{aff28}
\and
Department of Physics and Astronomy, University of New Mexico, Albuquerque, NM, 87131, USA\label{aff29}
\and
School of Physics and Astronomy, Cardiff University, Cardiff CF24 3YB, United Kingdom\label{aff30}
\and
Universit\'e Paris Cit\'e, CNRS, Astroparticule et Cosmologie, 75013 Paris, France\label{aff31}
\and
CNRS-UCB International Research Laboratory, Centre Pierre Bin\'etruy, IRL2007, CPB-IN2P3, Berkeley, USA\label{aff32}
\and
Dipartimento di Fisica - Sezione di Astronomia, Universit\`a di Trieste, Via Tiepolo 11, 34131 Trieste, Italy\label{aff33}
\and
Department of Physics, University of Cincinnati, Cincinnati,OH 45221, USA\label{aff34}
\and
ICSC - Centro Nazionale di Ricerca in High Performance Computing, Big Data e Quantum Computing, Via Magnanelli 2, Bologna, Italy\label{aff35}
\and
Dipartimento di Fisica, Universit\`a di Genova, Via Dodecaneso 33, 16146, Genova, Italy\label{aff36}
\and
INFN-Sezione di Genova, Via Dodecaneso 33, 16146, Genova, Italy\label{aff37}
\and
Department of Physics "E. Pancini", University Federico II, Via Cinthia 6, 80126, Napoli, Italy\label{aff38}
\and
INAF-Osservatorio Astronomico di Capodimonte, Via Moiariello 16, 80131 Napoli, Italy\label{aff39}
\and
Center for Astrophysics \textbar{} Harvard \& Smithsonian, 60 Garden Street, Cambridge, MA, 02138, USA\label{aff40}
\and
Dipartimento di Fisica, Universit\`a degli Studi di Torino, Via P. Giuria 1, 10125 Torino, Italy\label{aff41}
\and
INFN-Sezione di Torino, Via P. Giuria 1, 10125 Torino, Italy\label{aff42}
\and
INAF-Osservatorio Astrofisico di Torino, Via Osservatorio 20, 10025 Pino Torinese (TO), Italy\label{aff43}
\and
INAF-IASF Milano, Via Alfonso Corti 12, 20133 Milano, Italy\label{aff44}
\and
Enrico Fermi Institute, University of Chicago, 5640 South Ellis Avenue, Chicago, IL, 60637, USA\label{aff45}
\and
Department of Physics, University of Chicago, 5640 South Ellis Avenue, Chicago, IL, 60637, USA\label{aff46}
\and
Centro de Investigaciones Energ\'eticas, Medioambientales y Tecnol\'ogicas (CIEMAT), Avenida Complutense 40, 28040 Madrid, Spain\label{aff47}
\and
Port d'Informaci\'{o} Cient\'{i}fica, Campus UAB, C. Albareda s/n, 08193 Bellaterra (Barcelona), Spain\label{aff48}
\and
Istituto ricerche solari Aldo e Cele Dacc\`o (IRSOL), Faculty of Informatics, Universit\`a della Svizzera italiana, 6605 Locarno, Switzerland\label{aff49}
\and
Universit\'e de Gen\`eve, D\'epartement de Physique Th\'eorique, 24 Quai Ansermet, CH-1211 Gen\`eve 4, Switzerland\label{aff50}
\and
INAF-Osservatorio Astronomico di Roma, Via Frascati 33, 00078 Monteporzio Catone, Italy\label{aff51}
\and
INFN -- Sezione di Napoli, Via Cinthia 6, 80126, Napoli, Italy\label{aff52}
\and
Department of Physics and Astronomy, University of British Columbia, Vancouver, BC V6T 1Z1, Canada\label{aff53}
\and
National Research Council, Herzberg Astronomy and Astrophysics, 5071 West Saanich Rd., Victoria, V9E 2E7, Canada\label{aff54}
\and
Department of Physics and Atmospheric Science, Dalhousie University, Halifax, NS, B3H 4R2, Canada\label{aff55}
\and
National Taiwan University, No. 1, Sec. 4, Roosevelt Road, Taipei 106319, Taiwan\label{aff56}
\and
Dipartimento di Fisica e Astronomia "Augusto Righi" - Alma Mater Studiorum Universit\`a di Bologna, Viale Berti Pichat 6/2, 40127 Bologna, Italy\label{aff57}
\and
Department of Physics, University of California, Berkeley, CA, 94720, USA\label{aff58}
\and
Instituto de Astrof\'{\i}sica de Canarias, E-38205 La Laguna, Tenerife, Spain\label{aff59}
\and
Institute for Astronomy, University of Edinburgh, Royal Observatory, Blackford Hill, Edinburgh EH9 3HJ, UK\label{aff60}
\and
Jodrell Bank Centre for Astrophysics, Department of Physics and Astronomy, University of Manchester, Oxford Road, Manchester M13 9PL, UK\label{aff61}
\and
European Space Agency/ESRIN, Largo Galileo Galilei 1, 00044 Frascati, Roma, Italy\label{aff62}
\and
Universit\'e Lyon 1, CNRS/IN2P3, IP2I Lyon, Villeurbanne, France\label{aff63}
\and
Institut de Ci\`{e}ncies del Cosmos (ICCUB), Universitat de Barcelona (IEEC-UB), Mart\'{i} i Franqu\`{e}s 1, 08028 Barcelona, Spain\label{aff64}
\and
Instituci\'o Catalana de Recerca i Estudis Avan\c{c}ats (ICREA), Passeig de Llu\'{\i}s Companys 23, 08010 Barcelona, Spain\label{aff65}
\and
Institut de Ciencies de l'Espai (IEEC-CSIC), Campus UAB, Carrer de Can Magrans, s/n Cerdanyola del Vall\'es, 08193 Barcelona, Spain\label{aff66}
\and
Universit\'e Lyon 1, CNRS/IN2P3, IUF, IP2I Lyon, Villeurbanne, France\label{aff67}
\and
Mullard Space Science Laboratory, University College London, Holmbury St Mary, Dorking, Surrey RH5 6NT, UK\label{aff68}
\and
Universit\'e Paris-Saclay, Universit\'e Paris Cit\'e, CEA, CNRS, AIM, 91191, Gif-sur-Yvette, France\label{aff69}
\and
Department of Astronomy, University of Illinois Urbana-Champaign, 1002 West Green Street, Urbana, IL, 61801, USA\label{aff70}
\and
Universidad de La Laguna, Dpto. Astrof\'\i sica, E-38206 La Laguna, Tenerife, Spain\label{aff71}
\and
High Energy Accelerator Research Organization (KEK), Tsukuba, Ibaraki 305-0801, Japan\label{aff72}
\and
Inter-University Institute for Data Intensive Astronomy, Department of Astronomy, University of Cape Town, Cape Town, South Africa\label{aff73}
\and
Department of Physics, University of Pretoria, Hatfield, Pretoria, 0028, South Africa\label{aff74}
\and
Wits Centre for Astrophysics, University of the Witwatersrand, 1 Jan Smuts Avenue, 2000, Johannesburg, South Africa\label{aff75}
\and
Department of Astronomy, University of Geneva, ch. d'Ecogia 16, 1290 Versoix, Switzerland\label{aff76}
\and
Kapteyn Astronomical Institute, University of Groningen, Landleven 12, 9747 AD, Groningen, The Netherlands\label{aff77}
\and
Department of Physics and McGill Space Institute, McGill University, 3600 Rue University, Montreal, Quebec H3A 2T8, Canada\label{aff78}
\and
Canadian Institute for Advanced Research, CIFAR Program in Gravity and the Extreme Universe, Toronto, ON, M5G 1Z8, Canada\label{aff79}
\and
Universit\'e Paris-Saclay, CNRS, Institut d'astrophysique spatiale, 91405, Orsay, France\label{aff80}
\and
Institut universitaire de France (IUF), 1 rue Descartes, 75231 PARIS CEDEX 05, France\label{aff81}
\and
Joseph Henry Laboratories of Physics, Jadwin Hall, Princeton University, Princeton, NJ 08544, USA\label{aff82}
\and
Jet Propulsion Laboratory, California Institute of Technology, 4800 Oak Grove Drive, Pasadena, CA, 91109, USA\label{aff83}
\and
Max Planck Institute for Extraterrestrial Physics, Giessenbachstr. 1, 85748 Garching, Germany\label{aff84}
\and
Universit\"ats-Sternwarte M\"unchen, Fakult\"at f\"ur Physik, Ludwig-Maximilians-Universit\"at M\"unchen, Scheinerstr.~1, 81679 M\"unchen, Germany\label{aff85}
\and
INAF-Istituto di Astrofisica e Planetologia Spaziali, via del Fosso del Cavaliere, 100, 00100 Roma, Italy\label{aff86}
\and
Department of Astronomy, University of Science and Technology of China, Hefei 230026, China\label{aff87}
\and
School of Astronomy and Space Science, University of Science and Technology of China, Hefei 230026, China\label{aff88}
\and
Department of Physics, University of Illinois Urbana-Champaign, 1110 West Green Street, Urbana, IL, 61801, USA\label{aff89}
\and
Department of Physics and Astronomy, University of California, Los Angeles, CA, 90095, USA\label{aff90}
\and
Department of Physics and Astronomy, Michigan State University, East Lansing, MI 48824, USA\label{aff91}
\and
California Institute of Technology, 1200 East California Boulevard., Pasadena, CA, 91125, USA\label{aff92}
\and
INFN-Bologna, Via Irnerio 46, 40126 Bologna, Italy\label{aff93}
\and
Caltech/IPAC, 1200 E. California Blvd., Pasadena, CA 91125, USA\label{aff94}
\and
NSF NOIRLab, 950 N Cherry Ave, Tucson, AZ 85719, USA\label{aff95}
\and
Aix-Marseille Universit\'e, CNRS/IN2P3, CPPM, Marseille, France\label{aff96}
\and
Department of Physics, University of Oxford, Keble Road, Oxford OX1 3RH, UK\label{aff97}
\and
Instituto de F\'isica Te\'orica UAM-CSIC, Campus de Cantoblanco, 28049 Madrid, Spain\label{aff98}
\and
Department of Computer Science, Aalto University, PO Box 15400, Espoo, FI-00 076, Finland\label{aff99}
\and
Department of Physics, P.O. Box 64, University of Helsinki, 00014 Helsinki, Finland\label{aff100}
\and
Universit\"at Innsbruck, Institut f\"ur Astro- und Teilchenphysik, Technikerstr. 25/8, 6020 Innsbruck, Austria\label{aff101}
\and
INAF-Osservatorio Astronomico di Padova, Via dell'Osservatorio 5, 35122 Padova, Italy\label{aff102}
\and
Dipartimento di Fisica "Aldo Pontremoli", Universit\`a degli Studi di Milano, Via Celoria 16, 20133 Milano, Italy\label{aff103}
\and
INFN-Sezione di Milano, Via Celoria 16, 20133 Milano, Italy\label{aff104}
\and
CASA, Department of Astrophysical and Planetary Sciences, University of Colorado, Boulder, CO, 80309, USA\label{aff105}
\and
Department of Physics, University of Colorado, Boulder, CO, 80309, USA\label{aff106}
\and
Institute of Theoretical Astrophysics, University of Oslo, P.O. Box 1029 Blindern, 0315 Oslo, Norway\label{aff107}
\and
Leiden Observatory, Leiden University, Einsteinweg 55, 2333 CC Leiden, The Netherlands\label{aff108}
\and
Felix Hormuth Engineering, Goethestr. 17, 69181 Leimen, Germany\label{aff109}
\and
Technical University of Denmark, Elektrovej 327, 2800 Kgs. Lyngby, Denmark\label{aff110}
\and
Cosmic Dawn Center (DAWN), Denmark\label{aff111}
\and
Max-Planck-Institut f\"ur Astronomie, K\"onigstuhl 17, 69117 Heidelberg, Germany\label{aff112}
\and
Department of Physics, Centre for Extragalactic Astronomy, Durham University, South Road, Durham, DH1 3LE, UK\label{aff113}
\and
Department of Physics, Institute for Computational Cosmology, Durham University, South Road, Durham, DH1 3LE, UK\label{aff114}
\and
Astrophysics Research Centre, University of KwaZulu-Natal, Westville Campus, Durban 4041, South Africa\label{aff115}
\and
School of Mathematics, Statistics \& Computer Science, University of KwaZulu-Natal, Westville Campus, Durban 4041, South Africa\label{aff116}
\and
Institut de Recherche en Astrophysique et Plan\'etologie (IRAP), Universit\'e de Toulouse, CNRS, UPS, CNES, 14 Av. Edouard Belin, 31400 Toulouse, France\label{aff117}
\and
NASA Goddard Space Flight Center, Greenbelt, MD 20771, USA\label{aff118}
\and
Department of Astronomy \& the DiRAC Institute, University of Washington, Physics-Astronomy Building, Box 351580, Seattle, WA 98195-1700, USA\label{aff119}
\and
eScience Institute, University of Washington, Physics-Astronomy Building, Box 351580, Seattle, WA 98195-1700, USA\label{aff120}
\and
Universit\'e de Gen\`eve, D\'epartement de Physique Th\'eorique and Centre for Astroparticle Physics, 24 quai Ernest-Ansermet, CH-1211 Gen\`eve 4, Switzerland\label{aff121}
\and
Helsinki Institute of Physics, Gustaf H{\"a}llstr{\"o}min katu 2, University of Helsinki, 00014 Helsinki, Finland\label{aff122}
\and
Centre de Calcul de l'IN2P3/CNRS, 21 avenue Pierre de Coubertin 69627 Villeurbanne Cedex, France\label{aff123}
\and
Laboratoire d'etude de l'Univers et des phenomenes eXtremes, Observatoire de Paris, Universit\'e PSL, Sorbonne Universit\'e, CNRS, 92190 Meudon, France\label{aff124}
\and
Aix-Marseille Universit\'e, CNRS, CNES, LAM, Marseille, France\label{aff125}
\and
Department of Physics and Astronomy and George P. and Cynthia Woods Mitchell Institute for Fundamental Physics and Astronomy, Texas A\&M University, 4242 TAMU, College Station, TX 77843-4242, USA\label{aff126}
\and
SKAO, Jodrell Bank, Lower Withington, Macclesfield SK11 9FT, UK\label{aff127}
\and
Department of Physics \& Astronomy, Box 41051, Texas Tech University, Lubbock TX 79409-1051, USA\label{aff128}
\and
Univ. Grenoble Alpes, CNRS, Grenoble INP, LPSC-IN2P3, 53, Avenue des Martyrs, 38000 Grenoble, France\label{aff129}
\and
STAR Institute, Quartier Agora - All\'ee du six Ao\^ut, 19c B-4000 Li\'ege, Belgium\label{aff130}
\and
Laboratoire d’Annecy de Physique Theorique (LAPTh), CNRS/USMB, 9 Chemin de Bellevue BP110, Annecy, F-74941, ANNECY CEDEX, France\label{aff131}
\and
Universit\"at Bonn, Argelander-Institut f\"ur Astronomie, Auf dem H\"ugel 71, 53121 Bonn, Germany\label{aff132}
\and
Dipartimento di Fisica e Astronomia "Augusto Righi" - Alma Mater Studiorum Universit\`a di Bologna, via Piero Gobetti 93/2, 40129 Bologna, Italy\label{aff133}
\and
Institut f\"ur Theoretische Physik, University of Heidelberg, Philosophenweg 16, 69120 Heidelberg, Germany\label{aff134}
\and
Zentrum f\"ur Astronomie, Universit\"at Heidelberg, Philosophenweg 12, 69120 Heidelberg, Germany\label{aff135}
\and
School of Physics, HH Wills Physics Laboratory, University of Bristol, Tyndall Avenue, Bristol, BS8 1TL, UK\label{aff136}
\and
Universit\'e Paris-Saclay, CEA, D\'epartement de Physique des Particules, 91191, Gif-sur-Yvette, France\label{aff137}
\and
Prairie Research Institute, University of Illinois at Urbana-Champaign, 615 East Peabody Dr., Champaign, IL 61820, USA\label{aff138}
\and
Center for AstroPhysical Surveys, National Center for Supercomputing Applications, Urbana, IL, 61801, USA\label{aff139}
\and
Institute of Physics, Laboratory of Astrophysics, Ecole Polytechnique F\'ed\'erale de Lausanne (EPFL), Observatoire de Sauverny, 1290 Versoix, Switzerland\label{aff140}
\and
Telespazio UK S.L. for European Space Agency (ESA), Camino bajo del Castillo, s/n, Urbanizacion Villafranca del Castillo, Villanueva de la Ca\~nada, 28692 Madrid, Spain\label{aff141}
\and
School of Mathematics and Physics, University of Surrey, Guildford, Surrey, GU2 7XH, UK\label{aff142}
\and
European Space Agency/ESTEC, Keplerlaan 1, 2201 AZ Noordwijk, The Netherlands\label{aff143}
\and
School of Mathematics, Statistics and Physics, Newcastle University, Herschel Building, Newcastle-upon-Tyne, NE1 7RU, UK\label{aff144}
\and
NSF-Simons AI Institute for the Sky (SkAI), 172 E. Chestnut St., Chicago, IL 60611, USA\label{aff145}
\and
Cluster of Excellence "Our Dynamic Universe" (DYNAVERSE), Auf dem H\"ugel 71, 53121 Bonn, Germany\label{aff146}
\and
Institut de F\'{i}sica d'Altes Energies (IFAE), The Barcelona Institute of Science and Technology, Campus UAB, 08193 Bellaterra (Barcelona), Spain\label{aff147}
\and
Waterloo Centre for Astrophysics, University of Waterloo, Waterloo, Ontario N2L 3G1, Canada\label{aff148}
\and
Department of Physics and Astronomy, University of Waterloo, Waterloo, Ontario N2L 3G1, Canada\label{aff149}
\and
Perimeter Institute for Theoretical Physics, Waterloo, Ontario N2L 2Y5, Canada\label{aff150}
\and
Space Science Data Center, Italian Space Agency, via del Politecnico snc, 00133 Roma, Italy\label{aff151}
\and
Centre National d'Etudes Spatiales -- Centre spatial de Toulouse, 18 avenue Edouard Belin, 31401 Toulouse Cedex 9, France\label{aff152}
\and
Institute of Space Science, Str. Atomistilor, nr. 409 M\u{a}gurele, Ilfov, 077125, Romania\label{aff153}
\and
Dipartimento di Fisica e Astronomia "G. Galilei", Universit\`a di Padova, Via Marzolo 8, 35131 Padova, Italy\label{aff154}
\and
INFN-Padova, Via Marzolo 8, 35131 Padova, Italy\label{aff155}
\and
Department of Physics, Case Western Reserve University, Cleveland, OH, 44106, USA\label{aff156}
\and
Universit\'e St Joseph; Faculty of Sciences, Beirut, Lebanon\label{aff157}
\and
Departamento de F\'isica, FCFM, Universidad de Chile, Blanco Encalada 2008, Santiago, Chile\label{aff158}
\and
Department of Astronomy, University of Michigan, 1085 S. University Ave, Ann Arbor, MI 48109, USA\label{aff159}
\and
Department of Physics, Villanova University, 800 E Lancaster Ave, Villanova, PA 19085, USA\label{aff160}
\and
Department of Physics, Faculty of Science, Chulalongkorn University, 254 Phayathai Road, Pathumwan, Bangkok 10330, Thailand\label{aff161}
\and
Center for Astrophysics and Cosmology, University of Nova Gorica, Nova Gorica, Slovenia\label{aff162}
\and
CERCA/ISO, Department of Physics, Case Western Reserve University, Cleveland, OH 44106, USA\label{aff163}
\and
Departamento de F\'isica, Faculdade de Ci\^encias, Universidade de Lisboa, Edif\'icio C8, Campo Grande, PT1749-016 Lisboa, Portugal\label{aff164}
\and
Instituto de Astrof\'isica e Ci\^encias do Espa\c{c}o, Faculdade de Ci\^encias, Universidade de Lisboa, Campo Grande, 1749-016 Lisboa, Portugal\label{aff165}
\and
Cosmic Dawn Center (DAWN)\label{aff166}
\and
Niels Bohr Institute, University of Copenhagen, Jagtvej 128, 2200 Copenhagen, Denmark\label{aff167}
\and
Universidad Polit\'ecnica de Cartagena, Departamento de Electr\'onica y Tecnolog\'ia de Computadoras,  Plaza del Hospital 1, 30202 Cartagena, Spain\label{aff168}
\and
European University of Technology EUt+, European Union\label{aff169}
\and
Centre for Information Technology, University of Groningen, P.O. Box 11044, 9700 CA Groningen, The Netherlands\label{aff170}
\and
Institute of Space Sciences (ICE, CSIC), Campus UAB, Carrer de Can Magrans, s/n, 08193 Barcelona, Spain\label{aff171}
\and
Institut d'Estudis Espacials de Catalunya (IEEC),  Edifici RDIT, Campus UPC, 08860 Castelldefels, Barcelona, Spain\label{aff172}
\and
European Southern Observatory, Karl-Schwarzschild-Str. 2, 85748 Garching bei M\"{u}nchen, Germany\label{aff173}
}


 \abstract{We report the spectroscopic confirmation of a distant galaxy cluster, SPT-CL~J0111.8$-$5637.6, discovered through its thermal Sunyaev–Zeldovich (SZ) signature with a significance $\xi=10.6$ in the SPT-3G Main Field survey. 
 Euclid Data Release 1 imaging and photometric redshifts revealed a significant overdensity of red galaxies at a redshift $z\approx2.9$ that is coaligned with the SZ centroid. Spectroscopic observations with the Magellan/Baade telescope obtained  high-S/N detections of Ly\,$\alpha$ emission at $z=$ \zcluster \ for seven galaxies, including the likely brightest cluster galaxy. We used the redshift and SZ significance to infer a mass estimate of \msz, which makes SPT-CL~J0111.8$-$5637.6 the highest-redshift massive (\mass\, $> 1 \times 10^{14} M_\odot$) cluster yet detected.
  In addition to this extreme combination of mass and distance, there is a significant 
  extended dust emission feature spatially coincident with a secondary clump of red galaxies, the photometric redshifts of which are consistent with the cluster redshift. This system could be a direct link between the dusty star-forming era of protoclusters and the largely quiescent galaxy populations found in mature clusters.  }


   \maketitle

\nolinenumbers
\section{Introduction}

Distant massive galaxy clusters are invaluable for the study of cosmic structure formation \citep[e.g.,][]{allen11} and serve as astrophysical laboratories in which to probe the physical processes governing galaxy evolution \citep[e.g.,][]{kravtsov12}.
Numerous detailed studies of large cluster samples at $z<1$ suggest that the typical massive quiescent galaxy in the cores of these clusters formed half of its stars by $z\approx3$ \citep[e.g.,][and references therein]{Bower92,eisenhardt08, khullar22} and on reasonably short timescales \citep{Thomas05, Rettura10}.
Additional evidence for the importance of this early era comes from clusters at 1 $< z \lesssim 2$ that point to strong evolution in active galactic nucleus (AGN) activity and in star formation in cluster members as massive structures evolve from protoclusters \citep[e.g.,][and review by \citealp{Overzier16}]{overzier09, chiang13, wang16, miller18, daddi21, lepore24} to mature massive systems  \citep[e.g.,][and review by \citealp{alberts22}]{Stanford1998, brodwin13, Noirot16, vanderBurg20, mei23, trudeau24}. The high-redshift era at $z>2$ is thus of particular importance as it should directly sample still star-forming and active galaxies that later become the massive quiescent members of cluster cores.  

Despite the importance of the high-redshift era, observations have been sparse owing to the challenges in both finding and studying these distant rare objects. Existing high-redshift studies even at $z\gtrsim1.4$ are based on small samples composed of either infrared-selected clusters \citep[e.g.,][]{andreon09,stanford12,muzzin13}, rare X-ray detections \citep[e.g.][]{Mullis05,stanford06,hilton10, willis13}, galaxy overdensities in the environments of bright radio sources \citep{Wylezalek13, Wylezalek14, mei23}, or systems \citep[e.g.,][]{bleem15, khullar19, strazzullo19, old20, hilton26} from the first generation of blind surveys identifying clusters through the Sunyaev–Zeldovich effect \citep{sunyaev72}.

With the recent advent of a new generation of deep wide-field infrared \citep{akeson19, EuclidSkyOverview} and mm-wave surveys \citep{benson14, simons25} the landscape for high-redshift cluster discovery has fundamentally changed. 
Galaxy cluster detection via the thermal Sunyaev–Zeldovich (SZ) effect is a particularly powerful way to identify clusters out to the highest redshifts \citep{carlstrom02} via the signature their hot intracluster medium (ICM) imprints on the cosmic microwave background (CMB). 
The surface brightness of the signal is redshift-independent and has been found observationally in lower-redshift systems \citep[e.g.,][]{bocquet24, Umetsu26} and in simulations extending to higher redshifts \citep[e.g.,][]{Braspenning24} to be a low-scatter proxy for cluster mass. 
Optical and  near-infrared surveys, by contrast, detect clusters as significant galaxy overdensities in infrared colour \citep[e.g.,][]{Papovich10, muzzin13}, red-sequence \citep[e.g.,][]{gladders2000,rykoff14, klein26} or photometric redshift \citep[e.g.,][]{eisenhardt08,bellagamba18, thongkham24,wen24} space, providing complementarily selected samples as well as critical redshifts for the SZ detections. Weak lensing measurements from these surveys can also be used to constrain cluster mass-observable relations \citep[e.g.,][]{grandis24,EP-Ingoglia, finner26}. 

In this work, we combine observations from the SPT-3G camera on the South Pole
Telescope (SPT), data from the \textit{Euclid} survey, and follow-up
integral-field spectroscopy with the Large Lenslet Array Magellan Spectrograph (hereafter LLAMAS) on the Magellan telescope to confirm an extremely high-redshift massive SZ cluster, SPT-CL~J0111.8$-$5637.6, at $z=$ \zcluster. Our discussion of this system is organised as follows. In Sect.~\ref{sec:sz_discovery} we discuss the SZ detection in the SPT-3G Main Field survey and in Sect.~\ref{sec:euclid_confirmation}
its confirmation as a galaxy overdensity at $z\sim2.9$ in \textit{Euclid} near-infrared data. In  Sect.~\ref{sec:llamas} we describe the spectroscopic confirmation through detection of Ly\,$\alpha$ emission using Magellan/LLAMAS. In Sect.~\ref{sec:dust} we discuss another feature of the mm-wave detection: extended dust emission potentially connected to the cluster,
which may trace areas of high star formation in the cluster environment. Finally, we conclude in Sect.~\ref{sec:conclusion}. 

Conventions: Unless otherwise noted, we assume a fiducial flat $\Lambda$ cold dark matter (CDM) cosmology with $\sigma_8=0.80$, $\Omega_{\rm b} = 0.046$, $\Omega_{\rm m}~=~0.30$, $h = 0.70$, $n_{\rm s} = 0.972$ (at the pivot scale $k_{\rm s} = 0.002\,\mathrm{Mpc}^{-1}$), and $\Sigma m_\nu=0.06$\,eV. All magnitudes are reported in the AB system. Cluster masses are reported in terms of  $M_{200\mathrm{c}}$, which is defined as the mass enclosed within a radius, $r_{200\mathrm{c}}$, at which the average enclosed density is 200 times the critical density at the cluster redshift.

\section{SZ discovery of SPT-CL~J0111.8$-$5637.6}
\label{sec:sz_discovery}

SPT-CL~J0111.8$-$5637.6 (hereafter SPT0111) was discovered as part of an SZ cluster search in maps from five years of observations of the SPT-3G Main Field \citep[][hereafter B26]{bleem26}. This approximately 1600\,deg$^2$ field was surveyed to depths 
of 3.2, 2.5, and 8.9\,$\muk$-arcmin at 95, 150, and 220 GHz, respectively, as part of SPT-3G's campaign to obtain deep temperature and polarisation measurements of the CMB for cosmological \citep[e.g.,][]{camphuis26, omori26} and astrophysical \citep[e.g.,][]{guns26, simpson26} studies.  
Using multi-frequency matched filter techniques \citep[e.g.,][]{melin06},  B26 reported 8892 SZ cluster candidates detected above a significance $\xi=4$ in these data.  The quantity $\xi$ is SPT's primary SZ cluster observable and is defined as the highest signal-to-noise ratio (S/N) value recorded across all matched filter scales for a given SZ decrement. 
SPT0111 was detected as a high-significance candidate at $\xi=10.65$ with a preferred $\beta$-function profile with core radius $\theta_{\rm c}= \ang{;0.25;}$. 
Given SPT's arcmin-scale resolution and the high S/N of the detection, the candidate was well localised at Right Ascenion and Declination $(\alpha, \delta) = (\ang{17.9623}, \ang{-56.628}$), with a positional uncertainty of 7--9\arcsec \  which includes statistical uncertainty from the SPT cluster finder detection \citep{story11} and SPT's global astrometric uncertainty of 1--2\arcsec\  (Archipley et al. in prep.). 
Adopting the $\xi-$mass scaling relation of B26\footnote{The scaling relation is calibrated using an abundance matching analysis of the cluster sample at fixed cosmology, see section 5 in B26. The mass is Eddington bias corrected and the uncertainties include both intrinsic scatter and uncertainty on the redshift evolution. } and the  spectroscopic redshift of {$z=$ \zcluster} \ reported in Sect.~\ref{sec:llamas}, we estimate that SPT0111 has a mass \msz.

Analyses of archival imaging and catalogues from the Dark Energy Survey (DES) Y6
\citep{abbott21} and the Wide-field Infrared Survey Explorer \citep[WISE;][]{wright10,
meisner22}, together with dedicated Magellan/FourStar imaging \citep{persson13} deep
enough to confirm clusters up to $z\sim1.6$, found only weak evidence for an infrared
galaxy overdensity at the SZ position of SPT0111. These data were insufficient to provide a robust optical/infrared galaxy overdensity confirmation for the system. 

Above a significance of $\xi=6$, the SPT cluster candidate pool is essentially 100\% pure (see figure 4 in B26). 
For this reason, high-$\xi$ candidates without optical confirmation, and with no nearby stars or other low-redshift structures, were flagged as high-priority follow-up candidates.  Their high purity and lack of optical counterparts at $z \lesssim 1.4$--$1.6$ strongly suggested that these candidates are high-redshift systems.  In total, six systems at $\xi > 10$ and 28 systems at $\xi > 6$ were flagged, including SPT0111. 
As discussed further in Sect. \ref{sec:dust}, SPT0111 was also identified as a notable candidate by the presence of nearby extended 220\,GHz emission in the SPT-3G data, that is likely dominated by emission from dusty star-forming galaxies.  

\begin{figure*}
    \centering
    \includegraphics[width=0.49\linewidth]{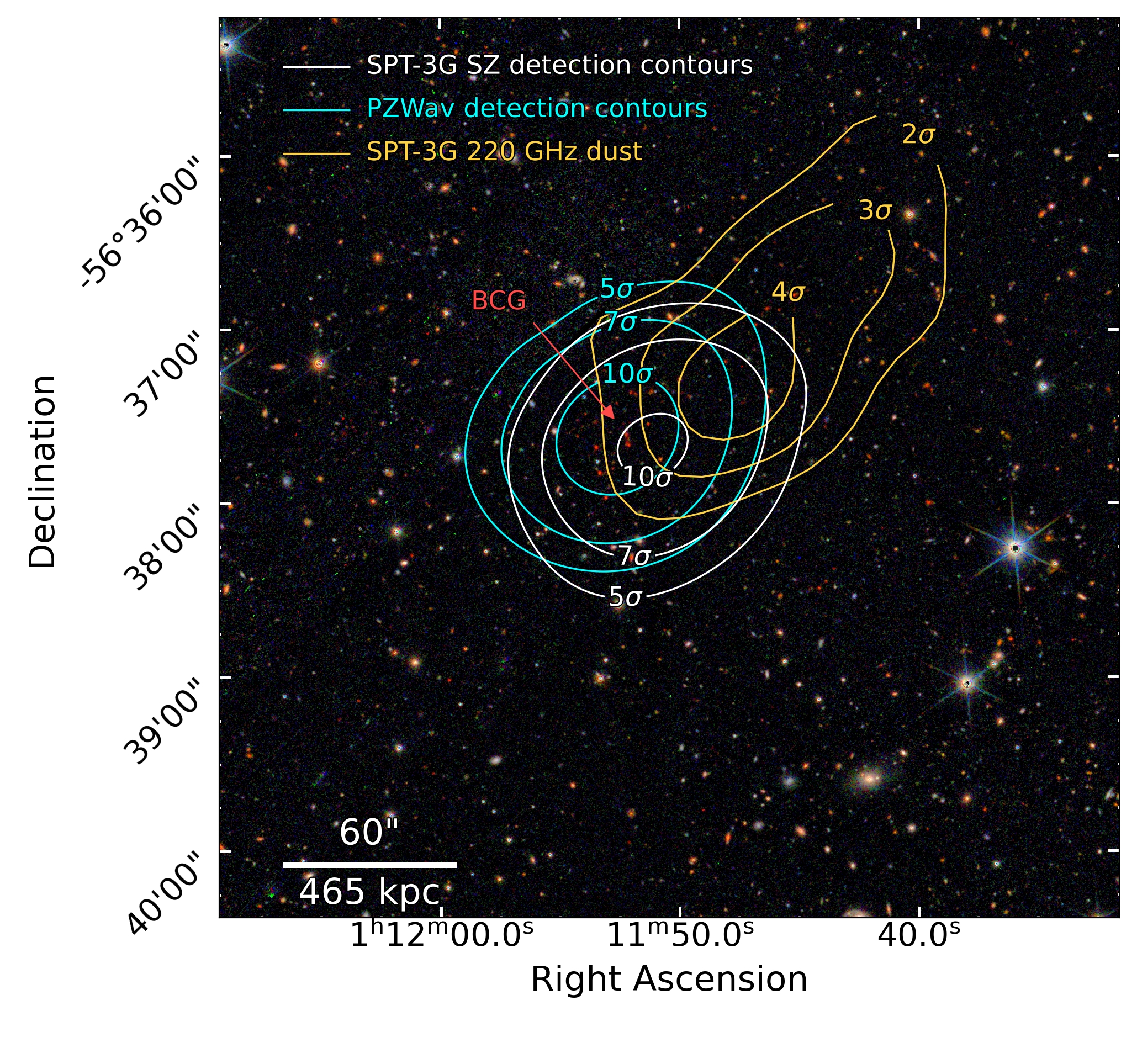}\hfill
    \includegraphics[width=0.49\linewidth]{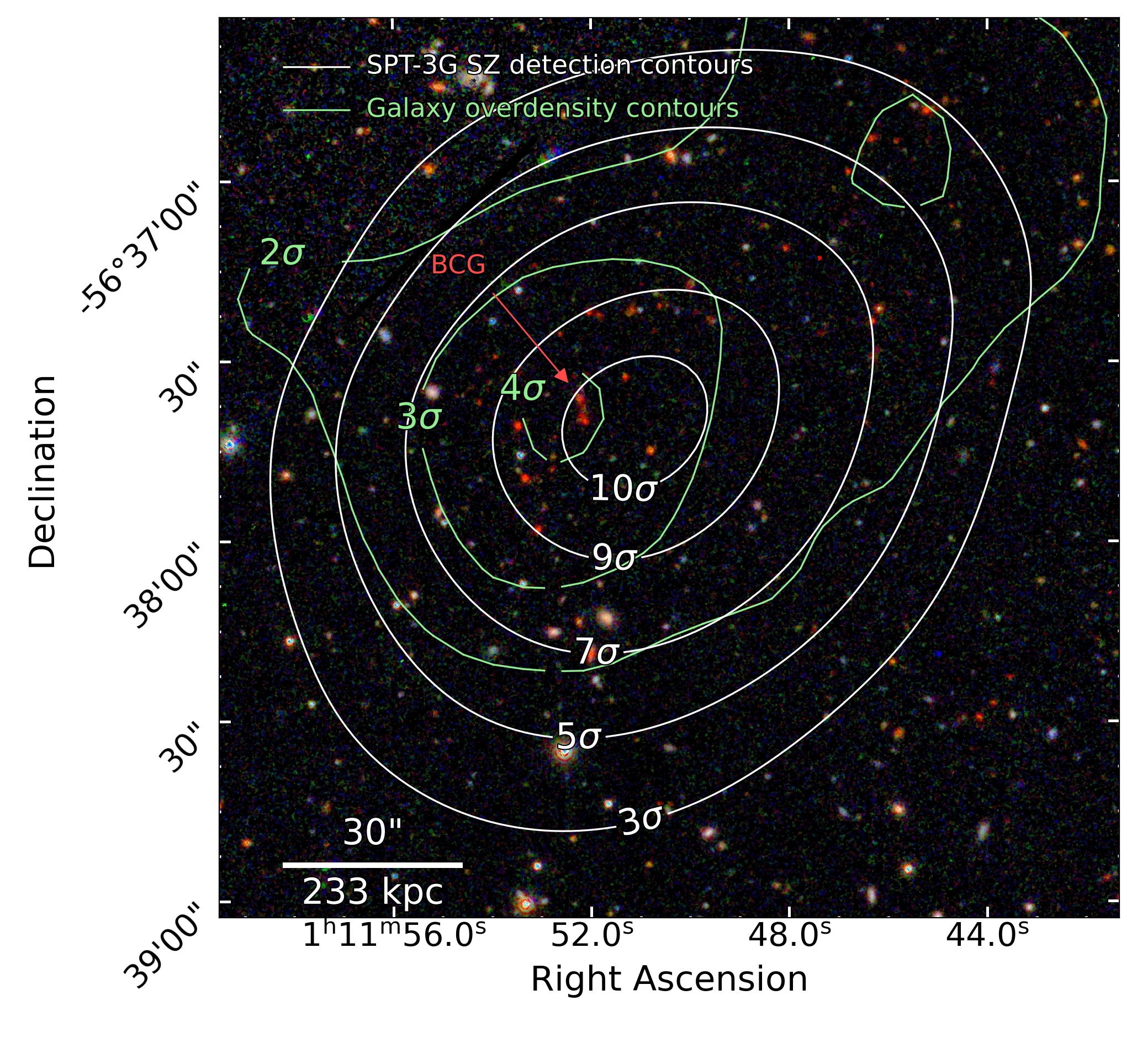}
    \vspace{-3mm}
    \caption{RGB colour images of SPT0111 made from the DR1 EWS \HE, \JE, and \YE imaging. \emph{Left}: This panel is \ang{;5.18;} across and centred on the SPT-3G position. Overlaid are the SPT-3G SZ significance contours (white), which peak at $\xi=10.6$; the \Euclid \texttt{PZWav} detection significance contours (cyan), which peak at $\mathrm{S/N}=$ 11.8; and contours corresponding to extended dust emission as traced by the SPT-3G 220\,GHz channel (orange, see Sect. \ref{sec:dust}).  For reference, the  full width at half maximum (FWHM) of the SPT 220~GHz beam is \ang{;1.04;}. \emph{Right}: This panel zooms in on the central \ang{;2.5;}, again showing the SZ significance contours. In this panel we also show light green galaxy density contours derived as described in Sect.~\ref{sec:photoz}. A pronounced overdensity of high-redshift red galaxies is evident, nearly coincident with the peak of the SPT contours. The red arrow indicates the location of LLAMAS Source 1 that we identify as the brightest cluster galaxy (see Sect.~\ref{sec:llamas}).}
    \label{fig:cluster_euclid}
    \vspace{-1mm}

\end{figure*}

\section{\Euclid confirmation and optical properties}
\label{sec:euclid_confirmation}

\textit{Euclid} is a European Space Agency mission currently conducting a 14\,000\,deg$^2$ optical and near infrared (NIR) imaging and spectroscopic survey of the extragalactic sky \citep{EuclidSkyOverview}. 
Observations from the visible imager \citep{EuclidSkyVIS} and from the Near-Infrared Spectrometer and Photometer 
\citep[NISP;][]{EuclidSkyNISP} provide coverage from 550 to 2020\,nm down to \IE, \YE, \JE, and \HE $=$ 26.0, 24.6, 24.6, and 24.5, respectively (5\,$\sigma$ point source sensitivity). 
These data are extremely valuable for the detection and confirmation of distant galaxy clusters \citep{Adam-EP3, Q1-SP050} with the expected recovery of cluster galaxies with high completeness to $m_*$+1.5 at redshifts $z\approx2$ and significant detection capabilities extending to even higher redshifts for rich clusters. 
Here $m_*$ is the apparent magnitude of a galaxy with the characteristic
luminosity $L_*$ of the \citet{schechter76} luminosity function.
Approximately 25\% of the SPT-3G Main Field is contained within the Data Release 1 Euclid Wide Survey (DR1 EWS; \citealt{DR1-TP001}).  

In this work, we use the photometric redshift (photo-$z$) catalogues produced 
for DR1 EWS \citep{DR1-TP012, DR1-TP011}.
The photo-$z$s were estimated using the Nearest-Neighbour Photometric Redshifts algorithm \citep[\texttt{NNPZ}; e.g.,][]{tanaka18,DR1-TP012}. 
This framework combines ground-based optical data (from DES in the SPT-3G footprint) with \textit{Euclid} photometry and then performs a nearest-neighbour match in colour--magnitude space to well-calibrated samples of reference galaxies. \texttt{NNPZ} returns multi-dimensional posterior distributions of both photo-$z$s  and galaxy physical properties. Detailed descriptions and validations of the \texttt{NNPZ} estimator and its application in the DR1 EWS can be found in \citet{DR1-TP012} and \citet{DR1-TP011}.

Examination of the unconfirmed high-$\xi$ SPT-3G cluster candidates in DR1 EWS data revealed a number of new high-$z$ cluster confirmations (which will be the subject of a forthcoming paper), and the highest-redshift of these systems is SPT0111. 
In Fig.~\ref{fig:cluster_euclid}, we show the \textit{Euclid} multi-band imaging in which there is a clear overdensity of faint red galaxies concentrated on the SZ location. 

\subsection{Photometric redshift analysis}
\label{sec:photoz}

The \Euclid cluster detection codes \texttt{PZWav} \citep[see][]{Gonzalez2014,thongkham24} and the Adaptive Matched Identifier of Cluster Objects \citep[hereafter \texttt{AMICO}; see][]{bellagamba18,Maturi2019}
were run on a 12\,$\deg^2$ field containing SPT0111. Both codes yielded  
strong detections nearly coincident with the SPT detection, with $\mathrm{S/N}=11.8$ at $z_{\rm phot}=2.94\pm0.08$ for \texttt{PZWav}, and $\mathrm{S/N}=18.6$ at $z_{\rm phot}=2.73$ for \texttt{AMICO} with comparable uncertainties.
Both photometric redshifts are consistent with one another despite the codes weighting galaxies differently for detection, and place SPT0111 at a redshift consistent with the spectroscopic measurement discussed in Sect.~\ref{sec:llamas}. This is true despite the expected limitations of the galaxy photo-$z$s at this epoch, owing to both the limited wavelength coverage of the \Euclid data above the 4000\,\AA\,break, and that the photo-$z$s are calibrated and validated at $z<2$. 

We used the \Euclid photometric redshifts 
for individual galaxies within 6 arcmin of the cluster centroid (i.e., 2.8\,Mpc at the cluster redshift) to derive a density map near the cluster core, adapting the methodology from \cite{gladders2000}. For each redshift interval of width 0.01, we convolved a fixed kernel with object sky positions. We follow the approach of the \textit{Euclid} cluster finders, weighting each object using a Gaussian probability distribution function centred at its median photometric redshift \texttt{\detokenize{PHZ_MEDIAN}} and with width defined by its 70\% photo-$z$ confidence interval \texttt{\detokenize{PHZ_INT_70}}.
We used a fixed kernel of the form $k(r)=A{\rm \,e}^{-1.965r}$, where $A$ is a normalisation constant and $r$ is in units of the Navarro--Frenk--White \citep[][NFW]{Navarro_1997}  scale radius -- chosen here, as in \cite{gladders2000}, to be 1/3\,Mpc 
at the slice redshift. 
With this method, we find a maximum galaxy overdensity at $z=2.8\pm0.3$.
In the right panel of Fig. \ref{fig:cluster_euclid}, the resulting density contours (shown  for $z=2.8$) highlight the presence of a secondary clump of probable cluster members northwest of the core. 
The locations of both the overdensities at the main cluster core and the secondary clump do not change significantly with redshift within the error bars.

\subsection{Red-sequence analysis}
We next consider whether there is already evidence for an established red sequence in this cluster. 
In particular, the brighter cluster galaxies appear very red in Fig.~\ref{fig:cluster_euclid}, suggesting either the presence of a population of passive red-sequence galaxies or of star-forming sources with substantial dust reddening. 
Passive galaxies are relatively rare in the field, particularly at high redshift ($z\approx3$), where they comprise only 4\% of the galaxy population at stellar masses of $10^{10}$ $M_\odot$ and reach 20\% at high stellar masses of $10^{11.5}$ $M_\odot$ \citep{weaver21}. In contrast, they often dominate the galaxy population in mature clusters. Overdensities of passive galaxies can therefore provide a strong contrast against the field population.

Building on our previous work on red-sequence-based cluster confirmation and redshift estimation using \Euclid data \citep{klein26}, we extend our multi-component matched filter \citep[MCMF;][]{klein19} method to $z\approx3$. For this purpose, we calibrated the red-sequence models beyond $z=2$ using $UVJ$-selected passive galaxies in the COSMOS field \citep{weaver21} with matched  \Euclid photometry. At lower redshifts, we use red-sequence models calibrated on clusters with known spectroscopic redshifts. We find good agreement between the two sets of models in the overlapping redshift range ($1<z<2$). Given limited statistics from COSMOS at redshifts $z>2$, our red-sequence models only contain estimates of the total scatter around the mean red-sequence models, while at lower redshifts magnitude dependent estimates of the intrinsic scatter are derived. 
At redshifts $z\gtrsim3$, the 4000\,\AA\,break of passive galaxies shifts into the \HE filter, imposing a natural limit on red-sequence-based techniques with the available filter set. 

Consistent with our previous work (B26), we find no significant intervening structure along the line of sight to SPT0111 out to $z\approx2.4$. 
As we move to higher redshift, we find a signal which increases and peaks towards the $z\approx3$ limit of the models, suggesting that the cluster redshift lies close to this limit, consistent with the photometric redshifts from the \Euclid cluster detection codes. 

\begin{figure}
    \centering
    \includegraphics[width=0.995\linewidth]{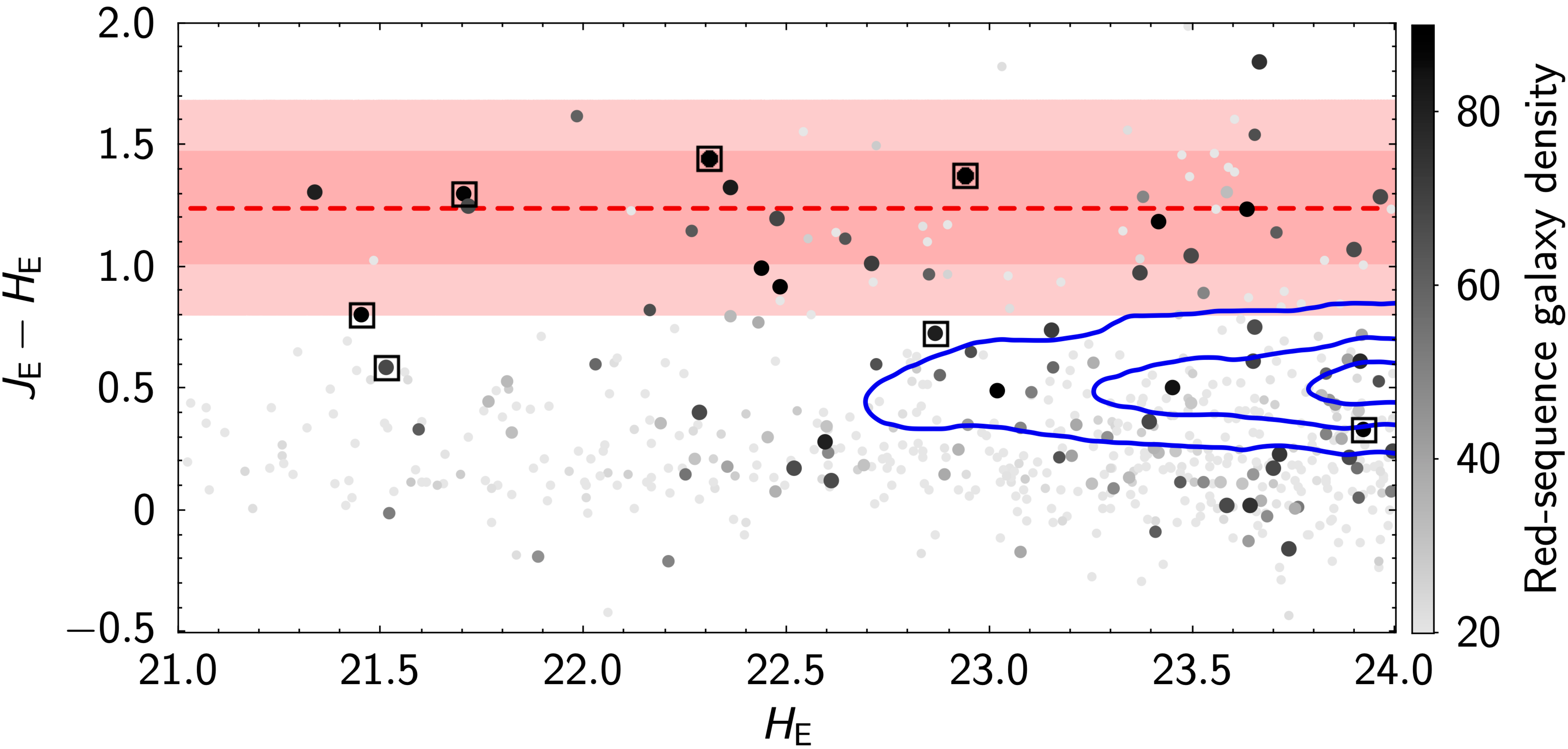}
    \includegraphics[width=0.90\linewidth]{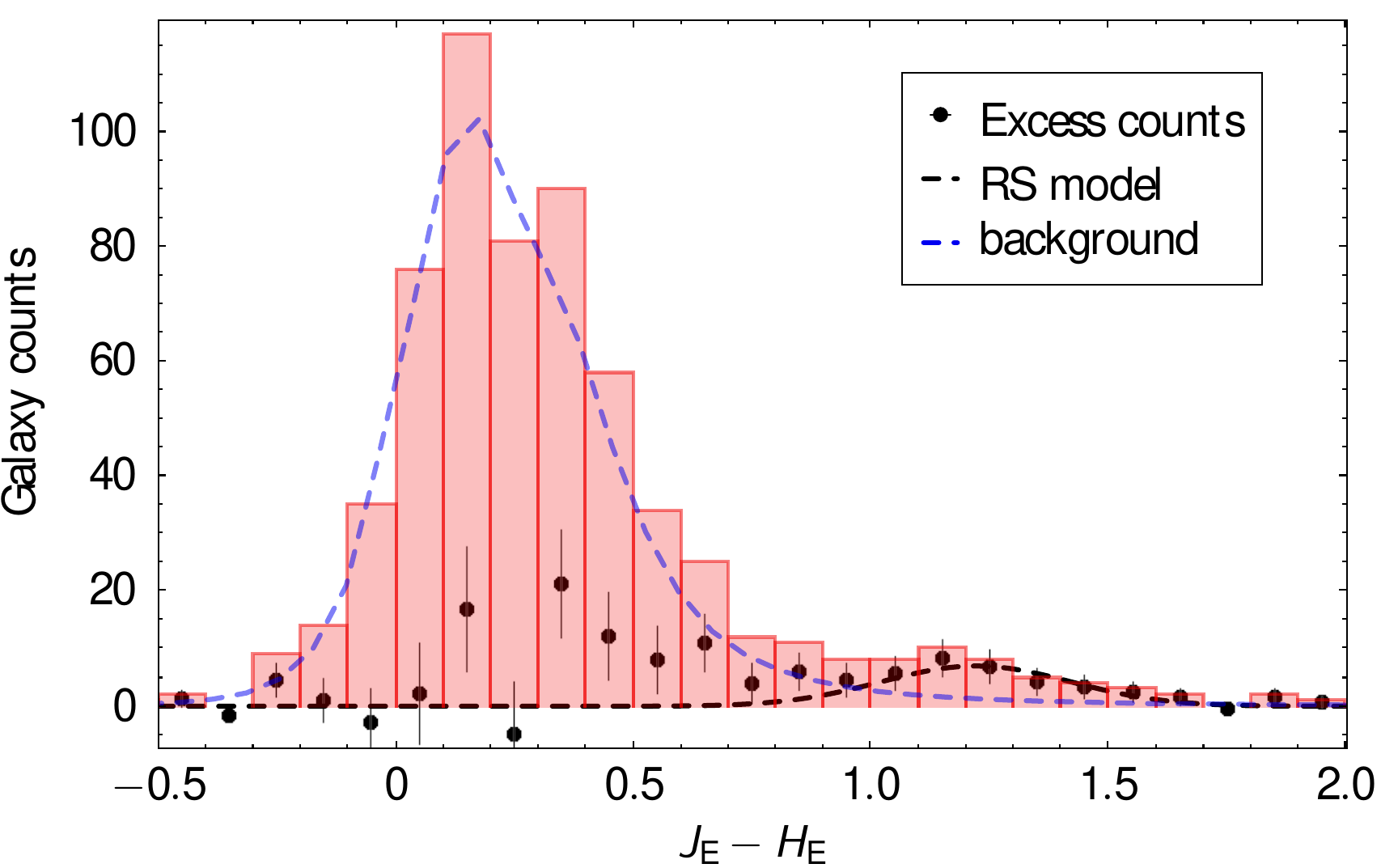}
    \caption{\emph{Top:} Colour--magnitude distribution of  \Euclid-detected galaxies within a 1\,Mpc ($125\arcsec$) radius around the SZ detection. The size and greyscale of the points is proportional to the red-sequence galaxy density, emphasising galaxies within the dense cluster core (i.e., larger and darker points indicate galaxies at higher local density). The red dashed line shows the expected average colour of red-sequence galaxies at $z=2.9$; shaded bands show 1 and $2\,\sigma$ scatter in the red-sequence model. Shown in blue are iso-density contours of star-forming galaxies in COSMOS at $2.8<z<3.0$, enclosing 80\%, 50\%, and 15\% of the star-forming galaxies. Black squares mark Ly\,$\alpha$ emitters from Table \ref{tab:lae}. \emph{Bottom:} Colour distribution of \Euclid-detected galaxies at $\HE < 24$ within a 1\,Mpc radius. The expected background distribution is shown as a blue dashed line, while the background subtracted colour distribution is shown as black points. A scaled normal distribution, with fixed variance and colour according to the red-sequence model at $z=2.9$, is shown by the black dash line.}
    \label{fig:cluster_euclid_CMD}
\end{figure}

To further investigate the high-redshift nature of the cluster, we examined the colour--magnitude distribution (CMD) of galaxies within 1\,Mpc ($\sim$125$\arcsec$ at $z=2.9$) of the SZ detection, with particular emphasis on the dense cluster core. In the top panel of Fig.~\ref{fig:cluster_euclid_CMD} we plot the CMD down to $\HE=24$, where the uncertainties on the $\JE - \HE$ colours are $\sigma_{\JE-\HE}<0.25$ for galaxies with colours consistent with the red-sequence, and in the bottom panel the $\JE - \HE$ colour distribution of the same galaxies. 
To increase the visibility of galaxies in the cluster core relative to projected field galaxies in the top panel, we scaled the intensity of each galaxy according to the red-sequence galaxy density at $z=2.9$, derived from galaxy density maps for galaxies within $2\,\sigma$ (0.43 mag) from the red-sequence colour. 
A joint analysis of $\JE - \HE$ and $\YE - \JE$ colours helps to constrain the redshift, suggesting a cluster redshift of  $z\approx2.9$. 

The spatially concentrated  brighter cluster galaxies have $\JE - \HE$ colours that are consistent with the expected location of the red sequence at $z=2.9$. This impression is further supported by the colour distribution of galaxies in excess of the background, shown in the bottom panel of Fig.~\ref{fig:cluster_euclid_CMD}, where the red part of the distribution is well described by a Gaussian with mean and variance expected from passive galaxies at $z=2.9$.
As seen from the blue contours in the top panel of Fig.~\ref{fig:cluster_euclid_CMD}, more than 80\% of the star-forming galaxies with $\HE<24$ at this redshift are expected to lie bluewards of the red-sequence model’s $2\,\sigma$ range. 
The overdensity of red-sequence-like galaxies within a 1\,Mpc radius is detected at $6.3\,\sigma$  
based on the \YE, \JE, \HE-band MCMF membership richness estimation down to $m_* + 2$ \citep{klein26} and measurements along \num{3200} random lines of sight. 

\begin{figure*}
    \centering
    \includegraphics[width=0.99\linewidth]{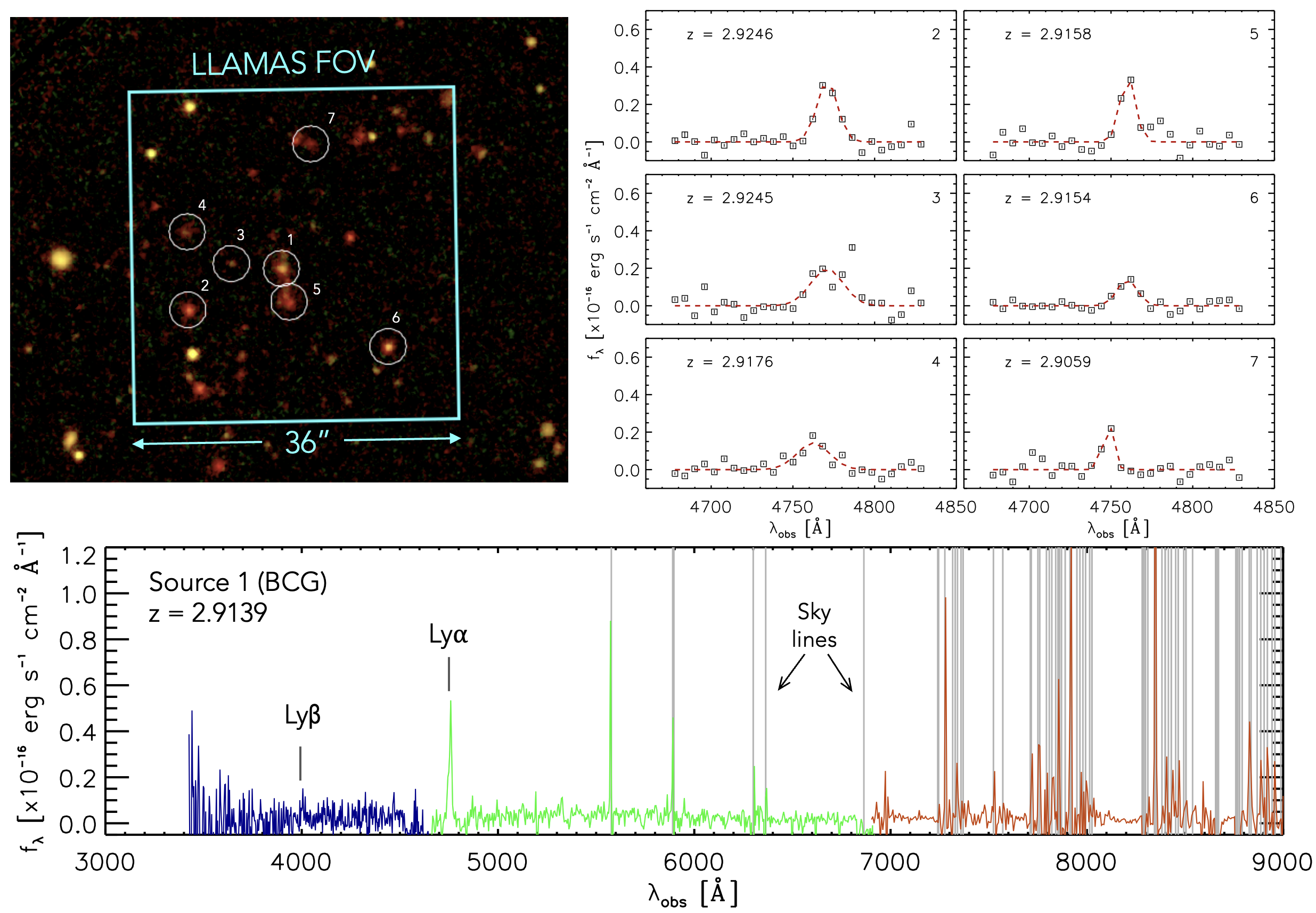}
    \caption{\emph{Top left}: Three-colour image based on \Euclid \IE-, \YE-, and \HE-band data. Overlaid on this image are the LLAMAS instrument field of view (cyan square) and the locations of seven Ly\,$\alpha$-emitting galaxies (white circles), identified via LLAMAS spectroscopy. \emph{Top right}: Spectra of the Ly\,$\alpha$ line extracted using a PSF-matched aperture at the locations of Sources 2--7, as labelled in the image on the left. Each of these sources exhibits a single bright emission line at $\sim$4760\,\AA, with an average redshift (assuming Ly\,$\alpha$) of $z=2.917 \pm 0.006$. \emph{Bottom}: Full LLAMAS spectrum (colour coded by spectrograph) for Source 1, which we refer to as a candidate brightest cluster galaxy (see Sect. \ref{sec:llamas}). The locations of bright sky lines are shown as vertical grey bands -- imperfect sky subtraction can lead to residuals at these wavelengths that look like emission lines. We label the expected locations of the Ly\,$\alpha$ and Ly~$\beta$ lines for a redshift of $z=2.9139$.  Over the full spectral range covered by the LLAMAS spectrograph, a single strong, high-significance emission line is detected, which is identified with high confidence as redshifted Ly\,$\alpha$.}
    \label{fig:llamas_spec_fig}
\end{figure*}

\subsection{\Euclid spectroscopy}
\label{sec:SPEZ}

Near-IR spectra from the Euclid Wide Survey (EWS), provided by the red grism (RGS) of NISP 
\citep[]{EuclidSkyNISP, DR1-TPSPE},
are available in the vicinity of SPT0111. Although the RGS spectroscopy could detect [O\,\textsc{ii}]$\lambda\lambda$3726,3729 and the rest-frame 4000\,\AA\,break in galaxies at $z \approx 2.9$, no such features were identified in any of the \Euclid spectra in the SPT0111 field.

\section{Spectroscopic confirmation with Magellan/LLAMAS}
\label{sec:llamas}

\begin{table*}
\centering
\caption{Ly\,$\alpha$ emitters in SPT0111. }  
\label{tab:lae}
\begin{tabular}{ccccc}
\hline
Source & RA & Dec & $z$ & $f_{\rm Ly\alpha}$\\
& [deg] & [deg] & & [$\times$10$^{-16}$ erg s$^{-1}$ cm$^{-2}$]\\
\hline
\noalign{\vskip 2pt}
1 & 17.9676 & $-$56.6267 & 2.9139 $\pm$ 0.0001 & 7.63 $\pm$ 0.16 \\
2 & 17.9728 & $-$56.6280 & 2.9246 $\pm$ 0.0001 & 5.10 $\pm$ 0.15 \\
3 & 17.9704 & $-$56.6266 & 2.9245 $\pm$ 0.0003 & 4.92 $\pm$ 0.14 \\
4 & 17.9729 & $-$56.6256 & 2.9176 $\pm$ 0.0004 & 3.56 $\pm$ 0.13 \\
5 & 17.9672 & $-$56.6277 & 2.9158 $\pm$ 0.0001 & 4.09 $\pm$ 0.11 \\
6 & 17.9616 & $-$56.6291 & 2.9154 $\pm$ 0.0003 & 2.13 $\pm$ 0.13 \\
7 & 17.9659 & $-$56.6229 & 2.9059 $\pm$ 0.0001 & 2.07 $\pm$ 0.08 \\
\hline
\end{tabular}
\par\smallskip
{\footnotesize \textit{Note.} Fluxes are measured in apertures of radius 1\farcs{25}.\par}

\end{table*}

SPT0111 was observed on 16 Oct 2026 (UT) with the LLAMAS\footnote{\url{https://sites.mit.edu/llamas/}} \citep{2020SPIE11447E..0AF} integral field spectrometer 
on the Magellan/Baade telescope. Seven $22.5$-min exposures were acquired, with small dithers in between, for a total integration time of 2.62 hours. 
The data were reduced using the \texttt{llamas-pyjamas} pipeline\footnote{\url{https://github.com/mit-kavli-institute/llamas-pyjamas}} (Hughes et al.\ in prep.), using a development branch built on v1.0.0 with improved fibre tracing, following repairs to two of the blue cameras in August 2026. 

The reduced and calibrated LLAMAS data cube spans observed frame 3500--9000\,\AA \ at a resolution R$\approx$2000--2200, and with a field of view of $36^{\prime\prime}\times\,36^{\prime\prime}$. 
We extracted spectra using apertures of radius 1\farcs{25} at the positions of the 21 \textit{Euclid}-detected galaxies at $\HE <24 $ within this field. 
Seven of these galaxies have a single emission line detected at $>3\,\sigma$ significance at 4750--4770\,\AA, consistent with Ly\,$\alpha$ at $z=2.917 \pm 0.006$ (see Fig. \ref{fig:llamas_spec_fig} and Table \ref{tab:lae}) as estimated using an unweighted mean. The typical flux uncertainty in these apertures is approximately 1$\times$10$^{-17}$ erg s$^{-1}$ cm$^{-2}$. 

We mark the Ly\,$\alpha$ emitters with black squares in the colour-magnitude diagram of Fig. \ref{fig:cluster_euclid_CMD}.
These seven sources all have red colours and are faint, consistent with being members of a high-redshift galaxy cluster, but in some cases are significantly bluer than the expected red-sequence.
This makes sense, since Ly\,$\alpha$ emission should be found primarily in or near star-forming galaxies \citep[e.g.,][]{Steidel2011}, rather than passive red-sequence galaxies. 
The fact that three of the Ly\,$\alpha$ emitters have colours consistent with the red sequence argues that they are  
potentially dusty, star-forming galaxies \citep{Chapman05}. Alternatively, these three could have substantial AGN components (note Source `2' is associated with a low-frequency radio source which is discussed briefly in Sect. \ref{sec:dust}), or even be associated with Ly\,$\alpha$ nebulae \citep[e.g.,][]{daddi21}. 

 We identify Source `1' (also marked in Fig. \ref{fig:cluster_euclid}), as a candidate brightest cluster galaxy (BCG) owing to its proximity to the SZ peak and its relative brightness both in Ly\,$\alpha$, in which it is the brightest emitter, and in the $\HE$ band, in which, at $\HE\sim21.4$, it is one of the brightest cluster members.
It additionally appears to be interacting with the second brightest \Euclid-detected galaxy in the LLAMAS FOV (labelled `5') and both are strongly detected in Ly\,$\alpha$ with a relative velocity of roughly 150\,km s$^{-1}$. 

The identification of the emission line as Ly\,$\alpha$ at {$z= $ \zcluster} \ merits further consideration.  There are alternatives such as C\,\textsc{iv}$\lambda$1549 at $z=2.07$, He\,\textsc{ii}$\lambda$1640 at $z=1.90$, C\,\textsc{iii}]$\lambda\lambda$1907,1909 at $z=1.49$, or [O\,\textsc{ii}]$\lambda\lambda$3726,3729 at $z=0.28$. We ruled out [O\,\textsc{ii}] at $z=0.28$ due to the simple fact that a group of seven galaxies at $z=0.28$ would be immediately obvious at the depth and resolution of \Euclid, and would show the entire suite of strong rest-optical lines expected at this redshift. There is no evidence for such low-redshift galaxies in the LLAMAS field of view.
The other possible rest-UV lines listed above are typically only seen in emission in AGN and in low-mass  galaxies \citep{bayliss14,stark14,Rigby15}, and only in those with  young and nearly dust-free stellar populations with typical UV continuum slopes of $\beta\approx-2$ and corresponding colours of $\approx0$, which is clearly not the case here.
The fact that a single bright line is detected in all seven sources over the wavelength range of 3500--9000\,\AA\ strongly suggests that this line is redshifted Ly\,$\alpha$, which is the brightest expected line blueward of 4760\,\AA\ at all redshifts and for all emission line sources. This is highlighted in the lower panel of Fig. \ref{fig:llamas_spec_fig}, which presents the full spectrum of the BCG, with a single high-significance detected emission line over the full wavelength range. We show in the lower panel the expected location of Ly~$\beta$, where there is a low-significance peak, providing further support for the measured redshift of $z=2.9139$ in the BCG.

Finally, the identification of the observed emission line as Ly\,$\alpha$ is strongly supported by the agreement between the implied spectroscopic redshift and the photometric results above, from both a full photo-$z$ analysis, and a red-sequence colour analysis. No other tenable solution exists: SPT0111 is spectroscopically confirmed at $z=2.92$.

\section{Correlated dust emission}
\label{sec:dust}

While the high-redshift SZ detection is remarkable on its own, SPT0111 is also flagged as an outlier in B26 in internal systematic checks of the recovered Compton-$y$ parameter designed to identify systems with either significant AGN or dust contamination.
Consistency checks between Compton-$y$ measured at 95~GHz versus 150~GHz alone find greater than $4\,\sigma$ evidence of dust-like contamination to the measured single-band SZ signal, with the 95~GHz estimate being higher than that obtained using 150~GHz data. Such strong contamination is rare. 
From the 8892 cluster candidates in B26, there are 35 candidates with similar $4\,\sigma$ discrepancies, and 159 candidates with $3\,\sigma$ dust-like discrepancies between the single-band $y$ measurements at 95 and 150 GHz.
For clusters at low redshift, where favourable geometry can lead to high gravitational lensing efficiencies for distant background sources \citep[see review of][]{Treu10}, such outliers can trace highly magnified dusty star-forming galaxies (DSFGs, e.g.,  \citealt{vanderlinde10,sun22}). At higher lens redshifts, where the angular diameter distance ratio in the lensing equation disfavours strong magnification (except for even rarer, very high-redshift background sources), this dust emission may be more likely to arise from emitters in the cluster environments \citep{kumar26}, although we cannot rule out chance line-of-sight projections from SPT data alone. 

	\begin{figure}[htbp]
		\centering
			\includegraphics[width=1\columnwidth]{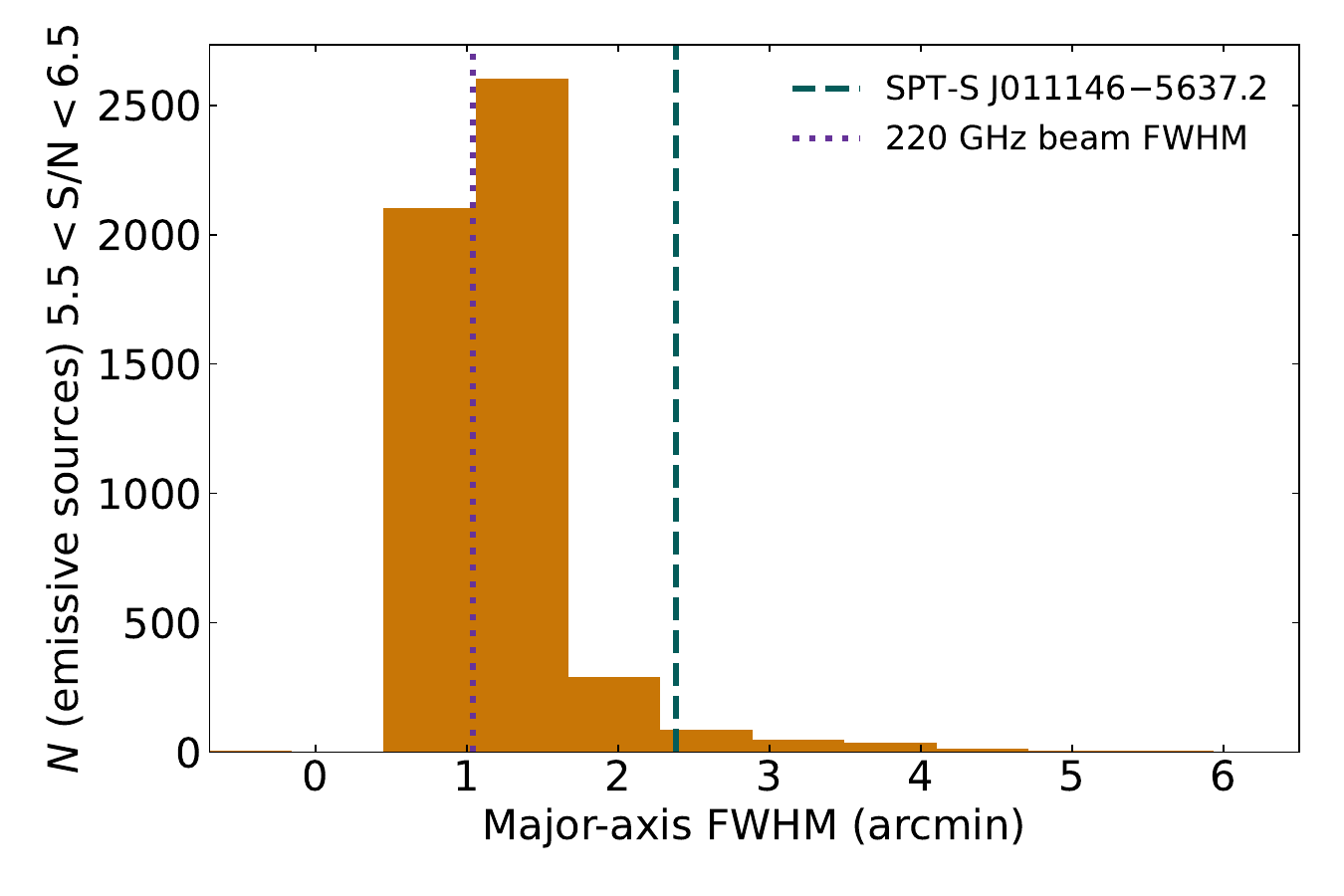}
			\caption{Histogram of major-axis FWHMs for emissive sources detected in SPT-3G 220 GHz data (Archipley et al. in prep.) at a similar significance to SPT-S J011146$-$5637.2, the dusty source close to SPT0111. This distribution shows that the source is likely to be extended compared to the 1\farcm{04} FWHM 220 GHz beam. The best-fit major-axis FWHM of 2\farcm{4} implies contribution to the emission from multiple galaxies.}
		\label{fig:emissivefwhms}
	\end{figure}

Examination of SPT-3G 220~GHz data (which, straddling the SZ null, contains negligible ICM signal) provides some plausibility for this second scenario. As reported in the forthcoming SPT-3G Main Field extragalactic source catalogue (Archipley et al. in prep.), we detect a $6.3\,\sigma$ dusty emissive source, SPT-S~J011146$-$5637.2, located $(42\pm6)\, {\rm arcsec}$ from the SZ decrement (orange contours in Fig. \ref{fig:cluster_euclid}). The dust centroid is offset approximately $4\,\sigma$ from the SZ centroid (accounting for the SZ positional uncertainty) and appears better aligned with a secondary peak of red galaxies. 
The 220~GHz emission is clearly elongated along the major axis of the galaxy overdensity, with a measured major-axis FWHM of 2\farcm{4}, significantly larger than the
220~GHz beam FWHM of 1\farcm{04} \citep{sobrin22}. 
At the cluster redshift, this emission would extend approximately 1.1~Mpc. 

We compared major-axis FWHM measurements of 220~GHz sources of similar detection S/N in Fig. \ref{fig:emissivefwhms}, and found that even at this relatively low detection significance, the measured extent is an outlier, indicating a rare resolved source. Such an extent suggests the contribution of multiple sources to the emission and also disfavours the lensing scenario where, based on the work of \citet{spilker16}, we would expect a lensed source to be unresolved in SPT data.  The total flux measured for this extended source is $(20.5\pm3.6)\,{\rm mJy}$ at 220\,GHz.

\subsection{Impact of dust emission on SZ properties}

The dust emission could partially fill in the cluster's recovered Compton-$y$ signal. 
Indeed, as noted above, the single band estimate at 150~GHz shows signs of this effect. 
By combining \textit{Herschel} and mm-wave data, \citet{Kornoelje2026} explore  the impact of such correlated dust emission on SZ cluster detection at $z\lesssim2$, finding that the combination of SPT-3G's deep three band data (with 95~GHz the least sensitive to dust and 220~GHz a strong dust tracer) alone was sufficient to mitigate bias.   
However, the emission associated with SPT0111 is significantly stronger: more than three times the median for clusters at $1.6\lesssim z\lesssim2$ in the SPT-3G sample of B26 (see fig.~16 in that work). We therefore perform several additional checks on the robustness of the SZ measurements.

First, SPT0111 is also detected in a blind cluster search in 95~GHz maps alone, with $\xi_{95}=8.3$ peaking at the $\theta_\mathrm{c}=0\farcm{25}$ filter scale. In 150~GHz data alone, it is detected at $\xi_{150}=2.8$ (as opposed to $\xi_{150}\approx5$ expected from noise estimates) because of the coincident dust emission. The recovered Compton-$y$ from 95~GHz data is within 1.5\% of the full multi-band combination, suggesting little contamination in the lowest frequency band.  
Second, we explored Compton-$y$ maps constructed through several different constrained internal linear combinations \citep[cILC;][]{remazeilles11} of the three band SPT-3G data, projecting out different plausible dust spectral energy distributions (SEDs). In the various combinations, the cluster is detected at a comparable S/N ($\xi=$ 9--11, noting that the cILC maps have marginally higher noise than the default multi-band maps), and that the SZ centroid shifts by only 4\arcsec, which is well within the statistical uncertainty.
These tests suggest that our derived SZ properties are robust to dust contamination. 

\noindent Radio Contamination: For completeness, we note that there is no evidence for strong mm-wave radio sources biasing the SZ detection of SPT0111. 
Australian Square Kilometre Array Pathfinder (ASKAP) 1.4~GHz catalogues \citep{duchesne24} contain three sources within 3 arcmin of the SZ centroid, with fluxes of 2.8, 2.2, and 246 mJy and offsets of 23\arcsec, 106\arcsec, \ and 138\arcsec, respectively. 
As is typical for low-frequency sources whose population is dominated by steep-spectrum sources \citep[e.g.,][]{dezotti10}, these objects are not strongly detected in SPT-3G data. Forced photometry of 95~GHz source detection maps revealed no appreciable signal for the closest source (which is also Source `2' in the Ly\,$\alpha$ emitters discussed in Sect. \ref{sec:llamas}) where the negative SZ decrement strongly dominates. 
The source offset by 106\arcsec \ is detected at $2.5\,\sigma$, and the one at 138\arcsec \ is detected at $3.7\,\sigma$ in SPT-3G data. 
This places the two larger-separation sources below the threshold for which filtering from the SPT map-making procedure could impart a positive bias greater than $1\,\sigma$ to the cluster detection (\citealp{Kornoelje2026}; see also \citealp{dicker24}).

\subsection{Potential origins of the dusty emission}
We now consider the physical origin of the dust emission if it is associated with the cluster. In the absence of lensing, single galaxies with 220 GHz flux density $S_{220} > 20$\,mJy are exceptionally rare \citep{Hill2018}, and thus we expect that the observed emission is the composite signal from multiple unresolved sources. With the current data we cannot distinguish whether this emission is due to a few very bright sources or a larger number of fainter sources, but sketch out two toy models to illustrate plausible physical scenarios. In both we assume that the combined flux density of all sources is $S_{220\mathrm{GHz}}=(20.5\pm3.6)\,{\rm mJy}$.

We first consider a case, Scenario 1, in which the dust emission arises from three equally bright, unlensed sources. Using the methods of \citet{Weiss2007}, \citet{Greve2012}, and \citet{Sulzenauer2026}, we estimated a dust mass of $M_\mathrm{d}=2.0_{-0.5}^{+0.6}\times10^9$\,$M_\odot$ per source ($6.0_{-1.5}^{+1.8}\times10^9$\,$M_\odot$ in total) -- implying (using a typical gas-to-dust conversion factor from e.g. \citealt{Saintonge2013}) a cold gas reservoir of $M_\mathrm{gas}\approx(6\mbox{--}12)\times10^{11}\,M_\odot$ -- and a star-formation rate (SFR) of $\mathrm{SFR}=2070_{-800}^{+1100}$\,$M_\odot$\,yr$^{-1}$ per source or $\mathrm{SFR}=6200_{-2300}^{+3300}\,M_\odot$\,yr$^{-1}$ in total. This SFR is comparable to that found in 220~GHz SPT-SZ selected protocluster cores \citep{Wang2021}.

Conversely, we consider Scenario 2, in which ten galaxies contribute equally to the extended dust emission. We find $M_\mathrm{d}=5.9_{-1.4}^{+1.9}\times10^8$\,$M_\odot$, $\mathrm{SFR}=620_{-220}^{+310}$\,$M_\odot$\,yr$^{-1}$, and $S_{870\mu\mathrm{m}}=(8.7\pm1.5)\,{\rm mJy}$. These properties are consistent with those of typical field DSFGs at $z \approx 2.7$ \citep{daCunha2015}. A group of ten DSFGs would, however, make this structure comparable to remarkable protocluster cores such as the Distant Red Core \citep{Oteo2018} or SPT2349$-$59 \citep{miller18}.

SPT0111 would be a remarkable system in either of these scenarios. In reality, the cluster likely hosts a mix of galaxies from these scenarios, along with contributions from fainter sources \citep[e.g.,][]{Kornoelje2026}. Clarifying the true origin of this dusty emission will require additional observations, with Atacama Large Millimeter/submillimeter Array (ALMA) data being particularly relevant as the spatial resolution would allow direct association with individual galaxies detected by \textit{Euclid}. In any case, the close projection of a massive dust reservoir approximately 1\,Mpc in projection from the SZ signal invites interpretations in the context of environmentally driven star-formation quenching. In such a scenario, the structure might have recently transitioned -- or be in the process of transitioning --  from a highly star-forming protocluster to a more quiescent cluster \citep{alberts22}, with vestigial dust reservoirs still present in the larger cluster environment.


\section{Discussion} 
\label{sec:conclusion}

At $z=2.917\pm0.006$ and \msz, SPT0111 is the most distant massive cluster ($\mass>10^{14}\,M_\odot$) discovered to date. 
It is among a small handful of spectroscopically confirmed ICM-detected groups and clusters at high ($z>1.8$) redshifts. 
These systems include JKCS~041 at $z= 1.803$ \citep{andreon09,andreon14,newman14}, with a weak lensing-calibrated mass of $\mass \approx 5 \times 10^{14}\ M_\odot$ \citep{kim25}, 
XLSSC 122 at $z=1.98$ \citep{willis13,willis20,mantz18} with a weak lensing-calibrated mass of $\mass =1.6 \times 10^{14}\ M_\odot$ \citep{Scofield26}, 
CL~J1449+0856 at $z=1.99$ with an XMM-\textit{Newton}- and \textit{Chandra}-derived mass estimate of \mass\, in the range (5.3--8) $\times\,10^{13}\ M_\odot$ \citep{gobat11,Gobat13}, 
CL~J1001+0220 \citep{wang16}  at $z=2.506$ 
with a \textit{Chandra}-derived mass estimate of $\mass \approx 8 \times 10^{13}\ M_\odot$ (although see \citealt{Champagne21} for a discussion of potential inverse Compton emission), 
and the group RO-1001 at $z=2.91$ with a $3.1\,\sigma$ \textit{Chandra} detection for a derived $\mass \sim 4\times 10^{13}\ M_\odot$ \citep{daddi21}. 
Several of these systems have also been detected in the SZ in deep high-resolution targeted observations with the Combined Array for Research in Millimeter-wave Astronomy \citep[XLSSC 122;][]{mantz14}, MUSTANG-2 \citep[JKCS~041;][]{Andreon23}, and ALMA (XLSSC 122; \citealt{vanMarrewijk24} and CL~J1449+0856; \citealt{gobat19}) as well as in wide field SZ cluster searches in the Atacama Cosmology Telescope DR6 data set (JKCS~041 at $\mathrm{S/N} =4.2$ and XLSSC 122 at $\mathrm{S/N} = 6.2$; \citealt{hilton26}). 

Among these systems SPT0111 stands out for its combination of high-mass and redshift, at roughly 5 times the mass of the closest ICM-detected system at a comparable redshift. 
We use the formalism of \citet{Harrison2013}, which was designed to address whether a single massive cluster at high redshift can falsify $\Lambda$CDM and quintessence models, to better assess the rarity of this cluster.
We  
calculated the likelihood of detecting such a cluster assuming a \citet{Castro-EP24} halo mass function\footnote{The results do not change qualitatively for a \citet{watson2013} halo mass function.} coupled with a \citet{Child2018} mass--concentration relation, and include uncertainty in evolution of the $\xi$--mass scaling relation. 

We find that the discovery of SPT0111 is slightly fortuitous but fully consistent with the concordance cosmological model.
We calculate 
a $97$\% likelihood of finding a cluster at this redshift or higher with at least this mass over the full sky. 
 Considering only the SPT-3G Main Field footprint ($\approx4$\% of the full sky),
the likelihood of finding a cluster with an observed mass as great as SPT0111 is 26.5\%. We note that the analysis above does not account for any systematic uncertainty arising from extrapolation of the $\xi-$mass relation to $z>2$. As a check, we also calculated the likelihoods using the weak-lensing-calibrated $\xi$-mass scaling relation from the cosmological analysis of \citet{bocquet24} assuming now a flat \citet{planck2018} $\Lambda$CDM cosmology, which yielded the same qualitative result. As for the expectations in the EWS, based on our  extrapolation of the forecast in \citet{Adam-EP3}, we expect $<$10 clusters as massive as SPT0111 at $z \approx 3$ in the full 14\,000 deg$^2$ EWS area.  
What we can confidently conclude at present is that this calculation confirms that SPT0111 is one of the most massive clusters expected to exist at this epoch. 

Both the Ly\,$\alpha$ detections and the unusual 220~GHz extended dust emission in close proximity to the SZ detection point to an active era for galaxies in the cluster environs. 
The detection of seven Ly\,$\alpha$-emitting galaxies in the inner approximately 100\,kpc of SPT0111 seems to suggest a significant amount of cold gas and star formation in this cluster, counter to what is observed in the nearby Universe. Ongoing follow-up studies of this system will assess in more detail the distribution and amount of multiphase gas in this unique system, as well as the detailed properties of the member galaxies. 
Additional high-resolution spatial and spectral follow-up observations are needed to ascertain the exact source of the dust emission. However, if verified to come from cluster members or infalling galaxies, this emission represents another bridge between massive clusters with a mature ICM and high redshift protoclusters with ICM detections such as the Spiderweb at $z=2.16$ \citep{dimascolo23,tozzi22,lepore24}, MQN01 at $z=3.25$ \citep{travascio26}, SPT2349–56 at $z=4.3$ \citep{chapman24,zhou2026}, and JADES-ID1 at $z \approx 5.7$ \citep{Bogdan26}.  

To conclude, in this paper we have presented the optical and spectroscopic confirmation of SPT0111, a recently discovered massive SZ cluster. 
While this exceptional system is the focus of a number of ongoing follow-up studies to better exploit its discovery for cosmological and astrophysical applications,  
we note with the advent of new SZ and NIR survey capabilities, we are rapidly transitioning away from an era in which we are limited by small numbers of confirmed very distant ($z>2$) clusters. 
The SPT-3G survey will detect $\sim100$ 
such systems at high S/N ($\xi>5$) in its 10\,000 deg$^2$ surveys \citep{raghunathan22}, while both the ongoing Simons Observatory \citep{simons25} and upcoming SPT-3G+ experiment \citep{natoli26} will each detect over 200 such systems.
The  \Euclid survey is forecast to detect $>100\,000$ clusters to $z=2$ \citep{sartoris16}, with--as demonstrated in this work--a significant number of higher-$z$ massive systems detected to $z=3$.  \textit{Roman} observations will further extend the NIR redshift reach \citep{Eifler21}.  
The combination of unprecedented wide-field NIR galaxy and SZ survey data promises to  be transformative for cluster science at the highest redshifts.

\begin{acknowledgements}
The South Pole Telescope program is supported by the National Science Foundation (NSF) through award OPP-2332483. Partial support is also provided by the Kavli Institute of Cosmological Physics at the University of Chicago. 
Argonne National Laboratory’s work was supported by the U.S. Department of Energy, Office of High Energy Physics, under contract DE-AC02-06CH11357. 
The UC Davis group acknowledges support from Michael and Ester Vaida. 
Work at the Fermi National Accelerator Laboratory (Fermilab), a U.S. Department of Energy, Office of Science, Office of High Energy Physics HEP User Facility, is managed by Fermi Forward Discovery Group, LLC, acting under Contract No. 89243024CSC000002.
The Melbourne authors acknowledge support from the  Australian Research Council’s Discovery Project scheme (No. DP260100705). 
The Paris group has received funding from the European Research Council (ERC) under the European Union’s Horizon 2020 research and innovation program (grant agreement No 101001897), and funding from the Centre National d’Etudes Spatiales. 
The SLAC group is supported in part by the Department of Energy at SLAC National Accelerator Laboratory, under contract DE-AC02-76SF00515.
This paper also uses data obtained with the 6.5-m Magellan Telescopes located at Las Campanas Observatory, Chile. Magellan observing time for this project was granted by the time allocation committees of the Massachusetts Institute of Technology. We would also like to express our gratitude to the staff of Las Campanas Observatory in Chile whose support was instrumental in facilitating data collection.
S.A.S, A.H.G, and D.S. acknowledge the support of NASA ROSES Grant 12-EUCLID11-0004. E.C. was supported by the National Science Foundation Graduate Research Fellowship Program under Grant No. 2140001. Any opinions, findings, and conclusions or recommendations expressed in this material are those of the author(s) and do not necessarily reflect the views of the National Science Foundation.
M. R. is supported by the Royal Society University Research Fellowship URF/R1/251742.
MA is supported by FONDECYT grant number 1252054, and gratefully acknowledges support from ANID Basal Project FB210003,  ANID MILENIO NCN2024\_112 and ANID + Vinculaci\'on Internacional + FOVI250261.
This work was supported in part by the U.S. Department of Energy under contract number DE-AC02-76SF00515.

\AckECol

\AckDRone

\end{acknowledgements}

\bibliographystyle{bibtex/aa}
\bibliography{bibtex/mind,Euclid.bib}

\begin{appendix}
  \onecolumn 

\end{appendix}

\end{document}